\documentclass[a4paper,UKenglish,cleveref, autoref,thm-restate]{lipics-v2021}

\hideLIPIcs

\usepackage{subcaption} 
\nolinenumbers
\usepackage{balance}
\usepackage{cite}
\usepackage{amsmath,amsfonts}
\usepackage{algcompatible}
\usepackage{graphicx}
\usepackage{textcomp}
\usepackage{xcolor}
\def\BibTeX{{\rm B\kern-.05em{\sc i\kern-.025em b}\kern-.08em
    T\kern-.1667em\lower.7ex\hbox{E}\kern-.125emX}}
\newcommand{\ignore}[1]{}

\newcommand{\squeezeup}{\vspace{-2.5mm}}

\usepackage[noend]{algpseudocode}

\usepackage{epsfig}
\usepackage{url}

\usepackage{ctable}

\definecolor{mygreen}{rgb}{0,0.5,0}
\definecolor{myred}{rgb}{0.8,0,0}
\definecolor{myblue}{rgb}{0,0,1}

\newcommand{\todo}[1]{\textcolor{red}{ToDo: #1}}

\usepackage{comment}

\newcommand{\hide}[1]{}

\usepackage{algpseudocode}
\usepackage{xpatch}

\usepackage{multicol}
\usepackage{algorithmicx}
\usepackage{algorithm}
\algtext*{EndIf}% Remove "end while" text
\algtext*{EndWhile}% Remove "end while" text
\algtext*{EndFunction}

\newcommand{\systemname}{\emph{Chopin}}

\title{Chopin: Combining Distributed and Centralized Schedulers for Self-Adjusting Datacenter Networks}

\author{Neta Rozen-Schiff}{School of computer science and engineering, Hebrew University, Israel}{neta.r.schiff@gmail.com}{https://orcid.org/0000-0003-1628-6843}{}

\author{Klaus-Tycho Foerster}{Computer Science Department, Technical University of Dortmund, Germany}{klaus-tycho.foerster@tu-dortmund.de}{https://orcid.org/0000-0003-4635-4480}{}

\author{Stefan Schmid}{TU Berlin, Germany\\ Faculty of Computer Science, University of Vienna, Austria}{stefan.schmid@tu-berlin.de}{https://orcid.org/0000-0002-7798-1711}{}

\author{David Hay}{School of computer science and engineering, Hebrew University, Israel}{dhay@cs.huji.ac.il }{https://orcid.org/0000-0001-9349-6049}{}

\titlerunning{Chopin: Combining Distributed and Centralized Schedulers}



\authorrunning{N. Rozen-Schiff, K.-T. Foerster, S. Schmid, and D. Hay}

\Copyright{N. Rozen-Schiff, K.-T. Foerster, S. Schmid, and D. Hay}

\funding{Research supported by the European Research Council (ERC), grant agreement No. 864228 (AdjustNet), Horizon 2020, 2020-2025, a grant from Fraunhofer SIT, and the Israeli Innovation Authority through the Peta-Cloud consortium.}

\supplementdetails[subcategory={Code}]{Software}{https://bitbucket.org/NetaRS/sched_analytics}

\begin{document}

\maketitle

%to shorten refs
%\bstctlcite{IEEEexample:BSTcontrol}

\begin{abstract}
The performance of distributed and data-centric applications often critically depends on the interconnecting network.
Emerging reconfigurable datacenter networks (RDCNs) are a particularly innovative approach to improve datacenter throughput. Relying on a dynamic optical topology which can be adjusted towards the workload in a demand-aware manner, RDCNs allow to exploit temporal and spatial locality in the communication pattern, and to provide topological shortcuts for frequently communicating racks. 
The key challenge, however, concerns how to realize demand-awareness in RDCNs in a scalable fashion. 

This paper presents and evaluates \systemname, a hybrid scheduler for self-adjusting networks that provides demand-awareness at low overhead, by combining centralized and distributed approaches. 
\systemname~allocates optical circuits to elephant flows, through its slower centralized scheduler, utilizing global information. 
\systemname's distributed scheduler is orders of magnitude faster and can swiftly react to changes in the traffic and adjust the optical circuits accordingly, by using only local information and running at each rack separately.

\end{abstract}

\vspace{2mm}
{\small \noindent\textbf{Bibliographical Note}: This paper will also appear in the Proceedings of the 26th International Conference on Principles of Distributed Systems (OPODIS 2022)~\cite{opodis2022}.}

\section{Introduction}\label{sec:intro}

Data-centric and distributed applications, including batch processing, streaming, scale-out databases, 
or distributed machine learning,
generate a significant amount of network traffic and their performance critically depends on the throughput of the underlying network~\cite{projectToR,RackCC}.

\begin{figure}[t]
%\vspace{-4mm}
\centering
\begin{subfigure}[t]{0.31\linewidth}
  \centering
  \includegraphics[%trim=0 0 1.0in 0, clip,
  width=\linewidth]{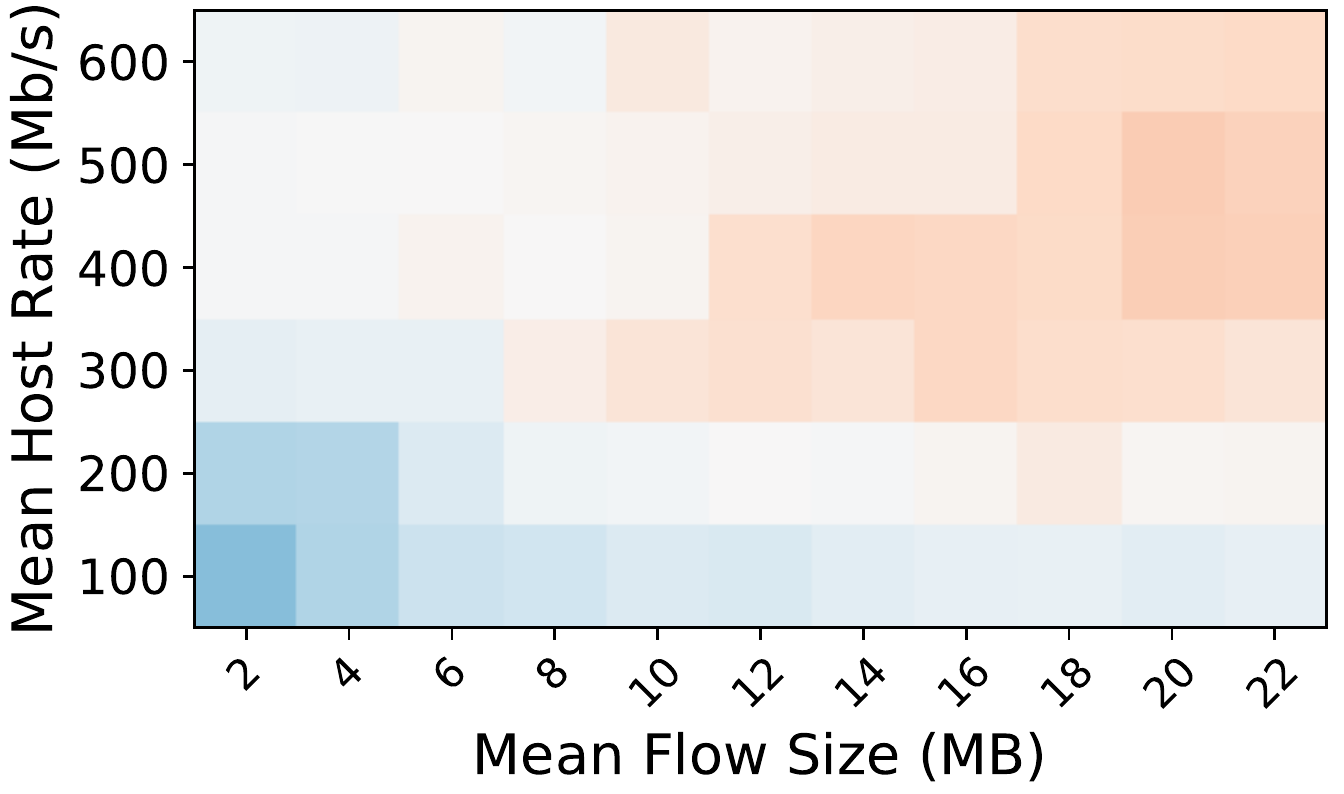}
  \caption{HULL~\cite{HULL}}
  \label{fig:mega_comb_heatmap_deg4_HULL}
  \hfill
\end{subfigure}%
\hfill
\begin{subfigure}[t]{0.31\linewidth}
  \centering
  \includegraphics[%trim=0 0 1.0in 0, clip,
  width=\linewidth]{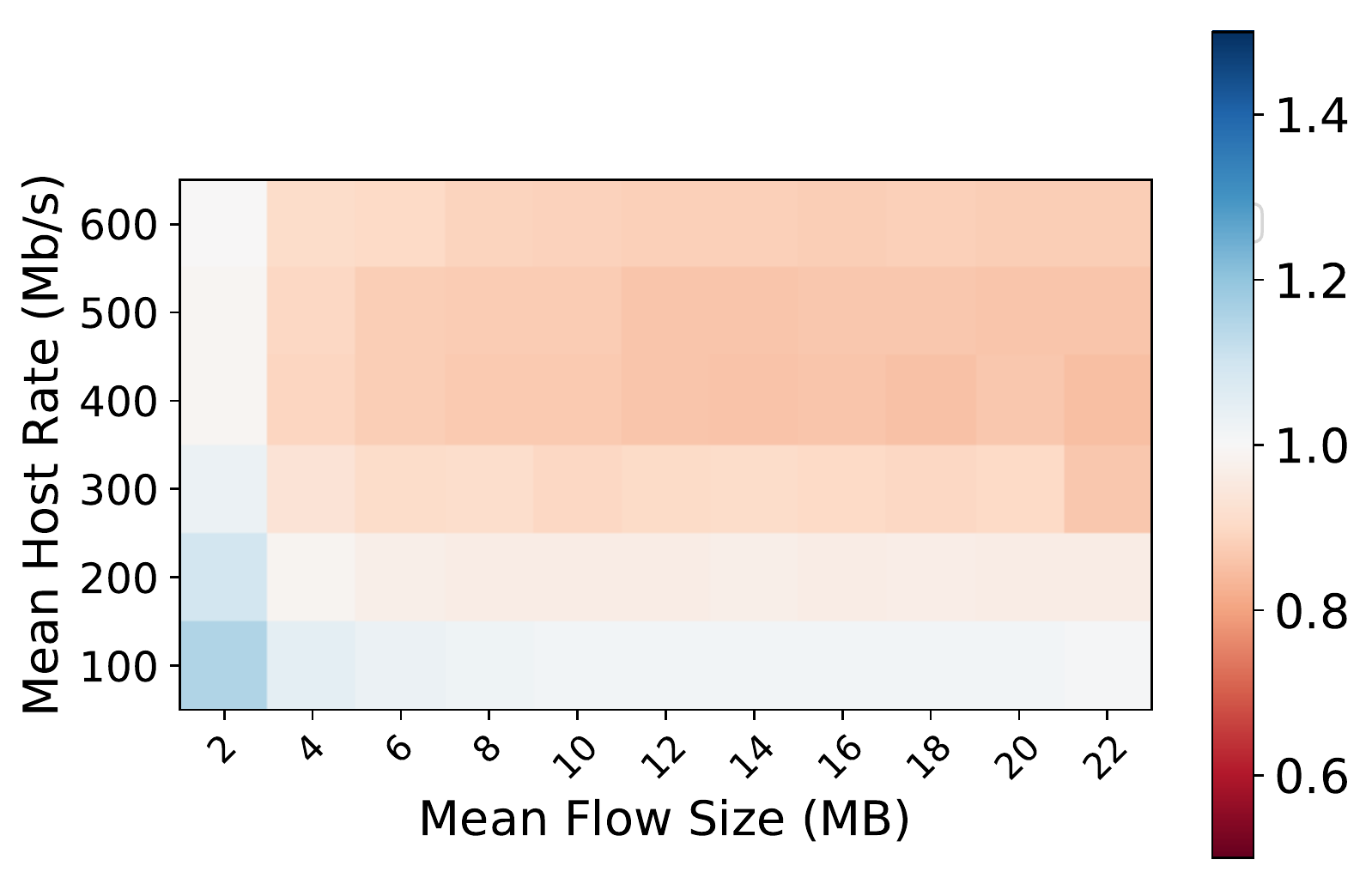}
  \caption{pFabric~\cite{DCTCP}}
  \label{fig:mega_comb_heatmap_deg4_pfabric}
  \hfill
\end{subfigure}%
\hfill
\begin{subfigure}[t]{0.31\linewidth}
  \centering
  \includegraphics[%trim=0 0 1.0in 0, clip, 
  width=\linewidth]{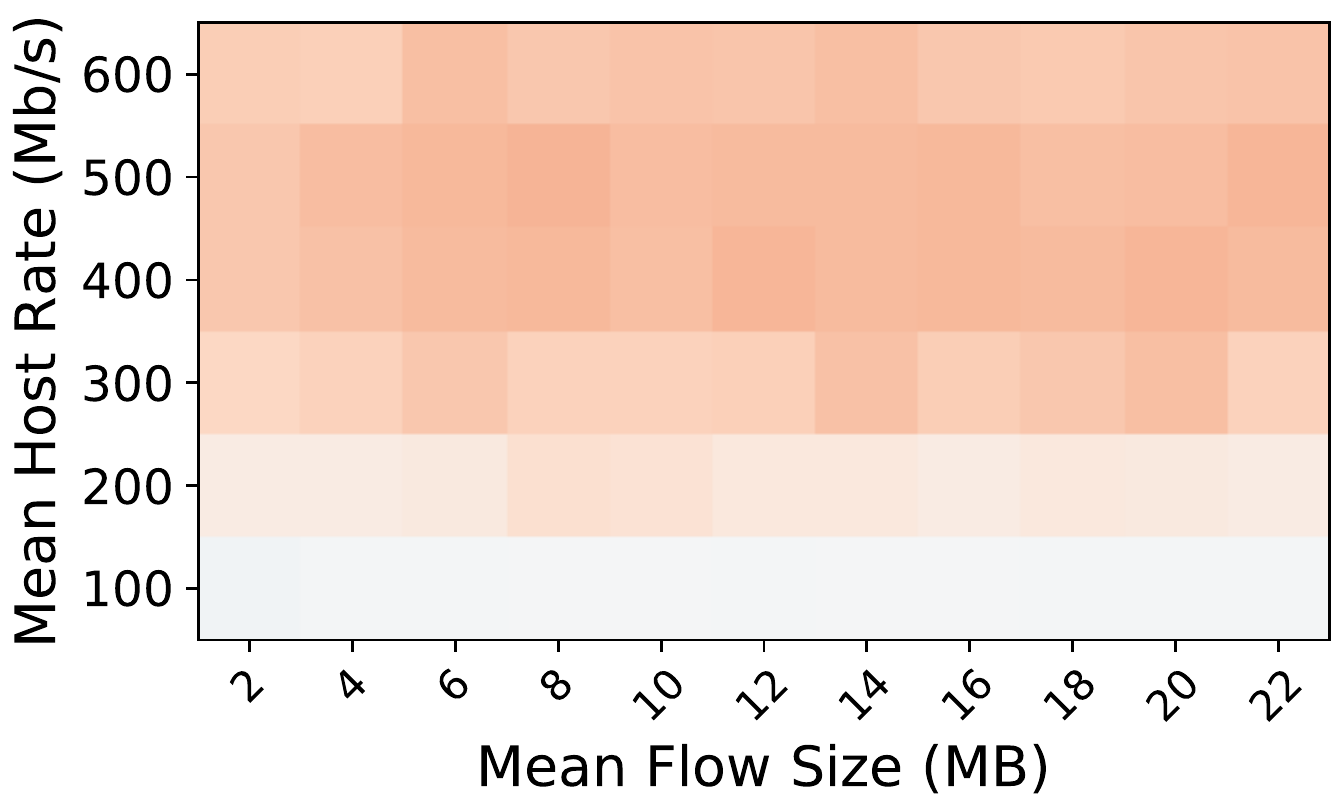}
  \caption{VL2~\cite{VL2}}
  \label{fig:mega_comb_heatmap_deg4_VL2}
  \hfill
\end{subfigure}%
\begin{subfigure}[t]{0.03\linewidth}
  \centering
  \includegraphics[%trim=0 0.5in 0 0, clip, width=\linewidth,
  scale=0.29]{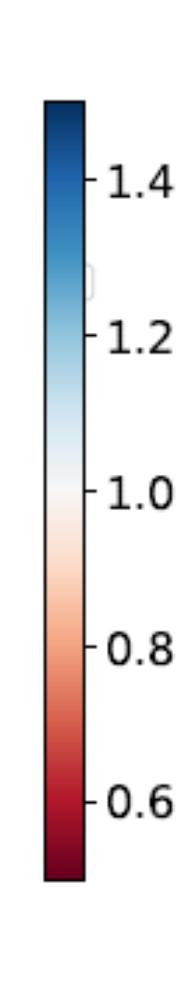}
\end{subfigure}%
%\vspace{-6mm}
\caption{Comparison between centralized and distributed schedulers under different traffic patterns (each generated by scaling well-known realistic flow size distributions, assuming Poisson flow arrival times under different rates), when the number of ToR switches is $80$, and the optical connectivity of each ToR switch is $4$. The color represents the ratio between the optical throughput of the distributed scheduler and the centralized scheduler. Blue cells mark settings where the distributed scheduler outperforms the centralized one; red cells mark the opposite. 
We refer to \S\ref{sec:Implementation} for topology details.
}
\label{fig:mega_comb_graph_deg4}
%\vspace{-5.5mm}
\end{figure}

%\subsection{Demand-aware Topologies Limitations}
To improve datacenter throughput, researchers and industry, e.g., Google~\cite{DBLP:conf/sigcomm/PoutievskiMOST022}, have recently started exploring innovative new datacenter designs that rely on dynamic and \emph{demand-aware} topologies: topologies that self-adjust toward the workload they currently serve.
The motivation behind self-adjusting datacenter topologies is twofold.

\emph{First}, empirical studies reveal that datacenter traffic patterns feature much structure~\cite{projectToR,sigmetrics20complexity,Facebook_2015,DCIntheWild}, i.e., are sparse, skewed, and bursty, which introduces optimization opportunities.
For example, a small number of flows typically carry the majority of traffic (these are called \emph{elephant flows}), while the remainder consists of a large number of flows that carry very little traffic (mice flows).
\ignore{For example, in a demand-aware network, elephant flows could be provided with very short routes, which reduces latency and network utilization, while improving throughput.}
Therefore, in a demand-aware network, the elephant flows should be routed through the optical circuits for offloading the electrical bottleneck, which in turn, reduces the latency of the mice flows, and improves the overall throughput.

\emph{Second}, emerging optical technologies and optical circuit switches enable the required very fast reconfigurations~\cite{Mordia, MordiaTimeAnalysis, MordiaToSDM, sirius}.
Over the last years, several interesting hybrid optical datacenter networks were suggested and evaluated~\cite{DBLP:conf/ancs/ZerwasA0B21}, augmenting an oversubscribed network with inter-rack optical links~\cite{projectToR,OSA,Helios,DBLP:journals/comcom/FenzFSV20,c_Through,FireFly,CircutSwitchUnderRadar,Megaswitch,Mordia,jellyfish,Solstice}, see~\cite{sigact19} for a survey. %google is ~\cite{DBLP:conf/sigcomm/PoutievskiMOST022}
The number of optical routes from/to each Top-of-Rack (ToR) switch, which we call the \emph{ToR switch optical degree}, is a single-digit number, typically at most $4$~\cite{MordiaTimeAnalysis, 4_connections, 4_port_8_port}.  

\vspace{1mm}
\noindent \textbf{Challenge: Scalability.}
While the vision of self-adjusting networks is intriguing and early solutions show promising results, the main challenge faced by such demand-aware networks concerns the scalability of the control plane.
Unlike demand-oblivious networks (i.e., static networks like Clos~\cite{clos1953study}, Slim Fly~\cite{slimfly}, and Xpander~\cite{xpander} or dynamic networks like RotorNet~\cite{RotorNet}, Opera~\cite{Opera} and Sirius~\cite{sirius}), 
demand-aware networks require the collection and evaluation of traffic patterns. 
In particular, performing all topology scheduling decisions \emph{centrally} (i.e., a centralized scheduler) may introduce a bottleneck and can result in slow reaction times. 
A fully distributed
%, i.e., \emph{local} and single-hop, 
decision making (i.e., a distributed scheduler) on the other hand,
may be suboptimal as it is based on incomplete information.

In order to show this tradeoff, we analyzed the \emph{optical throughput ratio}. The optical throughput ratio is defined as the ratio between the throughput routed through the optical circuit and the total datacenter throughput. It is a cornerstone measure as it reflects the utilization of the optical circuit, and therefore reduces the bottleneck over the electrical network.
Fig.~\ref{fig:mega_comb_graph_deg4} compares the  optical throughput ratio of the distributed-only scheduler and the centralized-only scheduler
%It demonstrates the tension between the two approaches, 
under different traffic patterns (each traffic pattern follows a distribution measured in a real datacenter, where we have parametrized the mean flow size and each host rate).
It demonstrates the tension between the two approaches: there is no ``clear winner'' and which one is better depends on the traffic pattern. The traffic pattern however is often not known when the datacenter is built and changes over time.  
For example, consider a datacenter serving a pFabric traffic pattern, with typical mean flow size of  approximately $1.7$ MB \cite{pFabric2}, and where each host sending rate is approximately $100$ Mbps. In this case, the optical circuit throughput ratio in the distributed scheduler is by $13\%$ higher compared to the centralized scheduler, as can be seen in the blue cells in Figure~\ref{fig:mega_comb_heatmap_deg4_pfabric}. However, for the same datacenter, with the same traffic pattern (pFabric), and the same mean flow size distribution, once the host's sending rate grows beyond $300$ Mbps, the centralized scheduler achieves a higher throughput ratio compared to the distributed one (see the relevant red~cell).

%We found that there is no clear hierarchy between distributed and centralized schedulers - for different traffic parameters (flow size distributions and host rates), the best among the two schedulers is different. 

Motivated by this insight, and by the desire to provide an efficient control plane for self-adjusting networks, we propose to combine both approaches to achieve the best 
of both worlds: 
fast reaction times of distributed decision-making and network utilization benefits of centralized optimization. 
 
 \vspace{1mm}

\noindent \textbf{Introducing Chopin.}
We present \systemname\footnote{Stands for: Controller for Hybrid OPtIcal electrical Networks.}, 
a novel scheduler for reconfigurable datacenter networks that fully exploits the benefits of self-adjusting networks by relying on an efficient control plane.
Specifically, \systemname\ provides demand-awareness at low overhead, by combining centralized and distributed approaches.
At the heart of Chopin's approach lies the idea that a relatively complex algorithm (e.g., Maximum Weight Matching, MWM) should be computed centrally, based on complete information.

However, since such an algorithm cannot be computed fast \cite{MWM_computation, MWM_computation_time, MWM_reconfiguration} (e.g., MWM may take around $20$ ms for $80$ ToR switches), we additionally allow distributed \emph{updates} to the centralized optical circuit allocation, based on a \emph{threshold}. 
The threshold specifies the flow weight changes from which a distributed scheduler can update the centralized scheduler allocation.
For example, if there is a large drop in demand in an allocated optical circuit (e.g., when an elephant flow ends), the distributed scheduler may tear it down and try to establish another circuit. % (based on local or single-hop information).
Hence, due to the volatility of many flows, we want a distributed constant-round algorithm (ideally just two rounds) and hence forgo more complex distributed algorithms~\cite{DBLP:journals/jacm/BalliuBHORS21} or dynamic centralized algorithms~\cite{DBLP:journals/algorithmica/BhattacharyaCH20}; the indirection via a centralized controller comes with overheads and delays which render this approach problematic to handle continuous update streams.

%where for the latter the control plane delay to a centralized controller is too large for a continuous update stream.

%\vspace{1mm}
\noindent \textbf{Our Contributions.}
In summary, we make the following contributions:
%\vspace{-1mm}
\begin{enumerate}
\item
We identify and analyze the difference in throughput performance of centralized and distributed schedulers for reconfigurable datacenter networks, for various scenarios and different flow size distributions. % To best of our knowledge, this is the first time such an analysis was made. 
\item
We design a hybrid scheduler, \systemname, which combines centralized and distributed decision-making based on thresholds.
To this end, we present and analyze both a centralized and a local online scheduler, exploring the trade-off between accuracy and running time. \systemname~relies on commodity devices available today, and required Chopin nodes which can simply be added to existing ToR switches by directing one of the switch ports to them. Moreover, information collection and dissemination of the centralized algorithm can be realized in the control plane using Software-Defined Networks (SDNs).
\item 
We report on Chopin's effectiveness through extensive simulations for different settings, showing that Chopin improves upon centralized and distributed approaches.
We achieve throughput improvements of up to $20\%$ against centralized and up to $23\%$ against distributed schedulers, \emph{always outperforming~both}.

%\todo{Chopin improve both centralized and distributed schedulers by approximately $20\%$.}

\end{enumerate}

\ignore{
\begin{figure}[t]
	%\vspace{-3mm}
 %\squeezeup
	\centering
	\includegraphics[width=0.42\textwidth]{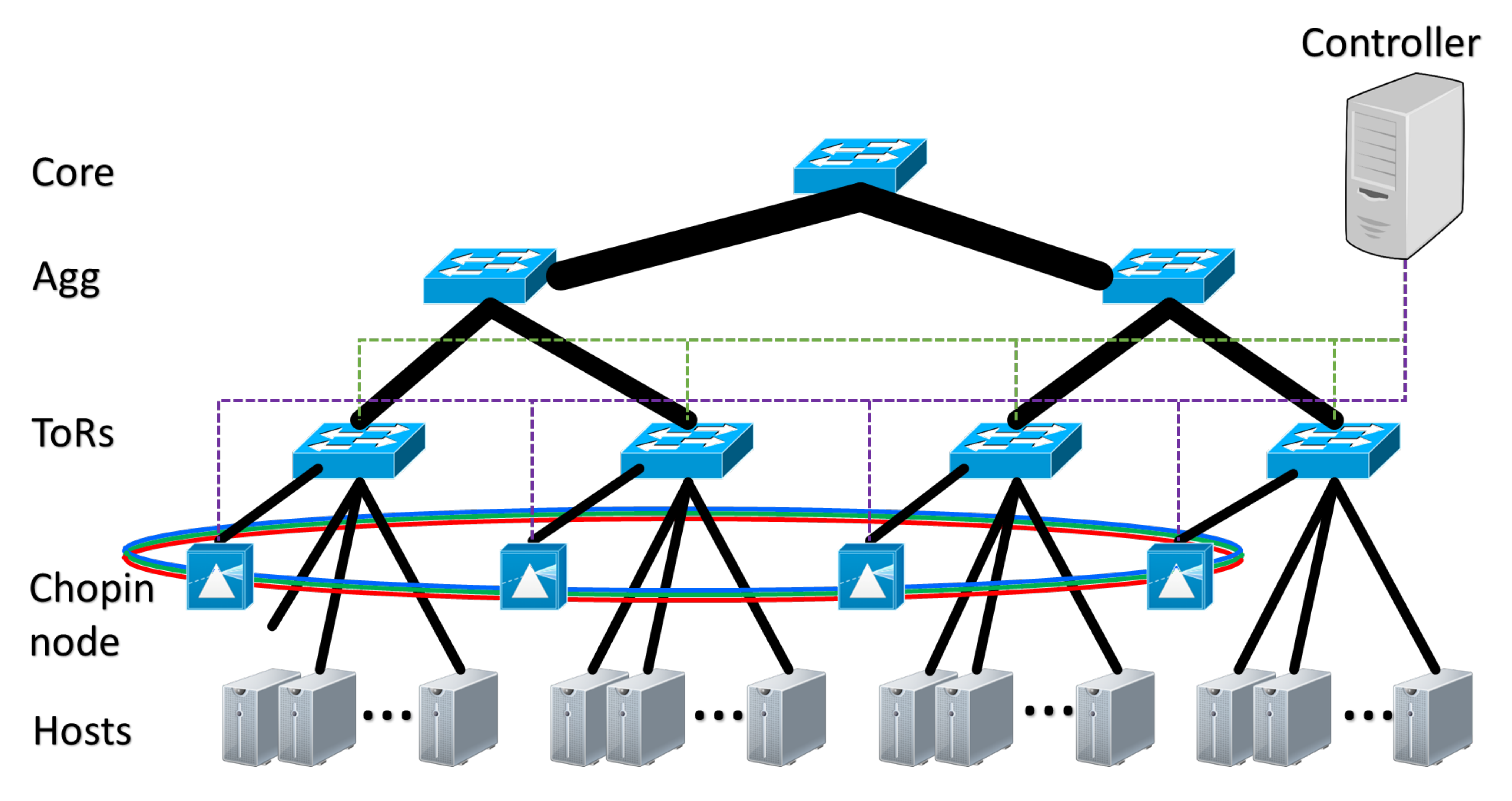}
	\vspace{-2mm}
	%\caption{Hierarchical datacenter equipped with Chopin circuits that interconnects its ToR switches.}
	\caption{Hierarchical DC with Chopin circuits interconnecting ToR switches.}
	\label{fig:Chopin_Concept_3d}
	 \squeezeup
  \vspace{-3mm}
\end{figure}
}

%\subsection{Organization}
%%We present the advantages and disadvantages of local and global schedulers in Section %\ref{sec:Central_Vs_distributed}. 
%We provide a high-level overview of the design of Chopin and formulation in Section~\ref{sec:design-chopin}. 
%%focusing on its hybrid topology, throughput optimizer. 
%The detailed design is divided into two schedulers: centralized and distributed. The centralized scheduler is %described in detail in Section~\ref{sec:centralized}, and  the distributed scheduler is presented in %Section~\ref{sec:distributed}. 
%We report on Chopin's simulations and evaluation in Section~\ref{sec:evaluation}. 
%Section~\ref{sec:Related_Work} reviews related works on optical switching and hybrid schedulers.
%Lastly, we conclude and discuss directions for future research in~Section~\ref{sec:conclusions}.
%\vspace{-2mm}

\section{Optical Background and Related Work}\label{sec:Related_Work}

Chopin is motivated by %builds upon the insights of previous work on hybrid datacenter networks, 
%in particular regarding 
trade-offs between centralized and distributed scheduling,\ignore{but also with respect to performance metrics and} which arise in matching algorithms. We first motivate why matching algorithms are central to Chopin's setting and then discuss centralized and distributed schedulers in this context.

\textbf{Optical Model: Why Matchings?} 
From a theoretical viewpoint, we consider the problem of how to augment a static network with (optical) edges in order to improve the total network performance. 
%
%This gives rise to the question: why should such augmentations be matchings?
The reason why this augmentation comes in the form of matchings lies in the 
%
%The reason lies in the
underlying hardware, namely optical circuit switches, we refer to Hall et al.~\cite[\S3]{DBLP:journals/osn/HallFSD21} for a technological overview.
In the simplest case, a set of nodes is connected to the optical circuit switch's ports by an optical cable each, and the switch ``matches'' these ports by e.g. adjusting mirrors to steer the light signals s.t. that pairs of ports (and hereby, pairs of nodes) are hence connected by optical circuits. 
Nodes could also be connected multiple times to the optical switch, or multiple optical switches could be used, giving rise to, e.g., $b$-matchings~\cite{DBLP:journals/ccr/FoersterPS19,DBLP:conf/infocom/HanauerHST22}.\footnote{There is also some work that considers multicast by splitting the outgoing light signals~\cite{DBLP:journals/jnca/LuoFSY22,DBLP:conf/icnp/SunN17,DBLP:journals/ton/DasRWWWCN22}.}
Conceptionally, other hardware could be used to the same effect (e.g., beamformed wireless connections~\cite{DBLP:conf/sigcomm/HalperinKPBW11} or free-space optics~\cite{FireFly}), but on a graph-theoretic level, they form circuits between pairs of nodes, and as thus, matchings.
We refer here to the survey by Foerster and Schmid~\cite{sigact19} for a further introduction to the enablers, algorithms, and complexity of reconfigurable datacenter networks.
We moreover refer to the article by Zerwas et al.~\cite{DBLP:conf/teletraffic/ZerwasKB21} on how system delays can be accounted for for scheduling algorithms.

\textbf{Centralized schedulers}
operate under the assumption of near-perfect utilization visibility and traffic demands, collected at a centralized location~\cite{Hawk}, often leveraging SDN. We refer to a recent survey and the references therein~\cite{DBLP:journals/comsur/ThyagaturuMMRK16}.
Herein the restriction to large and long-lived flows enables centralized schedulers~\cite{c_Through,Helios} to also cope with control loop delays.
However, these schedulers still suffer from traffic stability assumptions~\cite{huntingMice-TMS1}.\footnote{
Orthogonal to matching algorithms, Xia et al.~\cite{DBLP:conf/sigcomm/XiaSD0HN17} investigate how to migrate between Clos and random graph topologies. However, they require specialized $4/6$-port converter switches and also rely on a centralized control loop, estimating an update delay ``\textit{on the order of seconds}''~\cite{DBLP:conf/sigcomm/XiaSD0HN17}.
}
Traffic matrix schedulers~\cite{Mordia,REACToR,Solstice,eclipse} on the other hand, adjust packet transmissions to coincide with scheduled circuit reconfiguration, with full knowledge of when bandwidth will be available to particular destinations.
However, for the duration of the matching schedule, new flows are not accounted for and might need to wait for the next iteration.
In contrast, Chopin's design ensures rapid reactions to local traffic changes and new flow insertions, due to its additional distributed scheduler part.

\textbf{Distributed schedulers.}
In practice,  the large number of scheduling decisions and status reports can overwhelm centralized schedulers, and in turn lead to long latencies before scheduling
decisions are made~\cite{Hawk}.
ProjecToR~\cite{projectToR} initiated a broader interest in distributed scheduling, by proposing a stable-matching algorithm that optimizes for low latency, utilizing high fan-out single hop free-space optics~\cite{design-mirror}.
Via aging of requests, they obtain a constant-factor latency approximation for their online scheduling algorithm~\cite{devanur2016stable}.
RotorNet~\cite{RotorNet}, Opera~\cite{Opera}, and Sirius~\cite{sirius} employ a different approach and use lower fan-out circuits, where the topologies are created in a demand-oblivous manner.
RotorNet rotates through matchings independent of the current traffic, that provide eventual connectivity, where traffic is either scheduled to be routed along single hops, or along two hops, via buffering and a proposal and accept mechanism. %, with Opera using an expander-approach to always provide end-to-end connectivity.
Sirius follows similar ideas, either transmitting directly or via schemes reminiscent of Valiant's method.
Opera extends RotorNet by also always maintaining an expander graph, motivated by static topologies~\cite{xpander,DBLP:conf/sigcomm/KassingVSSS17}.
Although Opera's reconfiguration scheduling is deterministic, the precomputation of the topology layouts is in its current form still randomized.
Notwithstanding, ProjecToR, RotorNet, Sirius, and Opera can all rapidly deploy traffic along reconfigurable connections, by omitting a centralized control plane.
However, it is not clear how to realize the above three distributed systems with off-the-shelf hardware, such as a common optical circuit switch, and hence their application scenario is not as general as with Chopin.
Notwithstanding, Decentralized scheduling is also used in several other systems, including SplayNet~\cite{schmid2015splaynet}, Cerberus~\cite{sigmetrics22cerberus}, or CacheNet~\cite{comnet22}.

Lastly, while there is profound research on matching algorithms in the distributed computing community~\cite{DBLP:journals/csur/Suomela13}, distributed algorithms for maximal matchings in graphs with large degree $\Delta$ (as for optical circuit switches) are relatively slow~\cite{DBLP:journals/jacm/BalliuBHORS21}.
While approximation~\cite{DBLP:journals/jacm/LotkerPP15} and dynamic~\cite{DBLP:journals/siamcomp/LotkerPR09}  algorithms are considerably faster, here the constraints of the optical datacenter networking and the distributed computing community are quite different and hence the communities (yet) don't overlap much in their research applications: ideally, for optical circuit switching, small-constant round algorithms of low computational complexity are desired, whereas in the distributed computing community, the local algorithms can be more complex, with a focus on asymptotic runtime optimization.
As thus, Chopin utilizes a low complexity threshold based distributed algorithm, using just two rounds of communication, which falls in line with the requirements of hybrid datacenters.

%
%\klaus{klaus: also discuss non-optical, but highlight that it is non-circuit: \url{https://engineering.fb.com/data-center-engineering/building-express-backbone-facebook-s-new-long-haul-network/}~\cite{fbcd}}
%
%\newline\indent
%
%Decentralized scheduling is also used in several other systems,  including SplayNet~\cite{schmid2015splaynet}, Cerberus~\cite{sigmetrics22cerberus}, or CacheNet~\cite{comnet22}. 
%However, completely distributed algorithms can make poor scheduling decisions because of limited visibility into the overall resource usage in the cluster \cite{Hawk}.

\section{Chopin's Design}\label{sec:design-chopin}

%We start by providing an overview of Chopin's design.
%More details will follow in subsequent sections. 

In a nutshell, \systemname's topology scheduler aims to provide demand-awareness efficiently by combining centrally optimized decision making with fast distributed reactions. 
The idea is hence analogous to the nervous system of animals, which is typically divided into a slower central nervous system and a faster peripheral nervous system~\cite{tortora2018principles}.

\begin{figure}[b]
	\centering
	\includegraphics[width=0.60\textwidth]{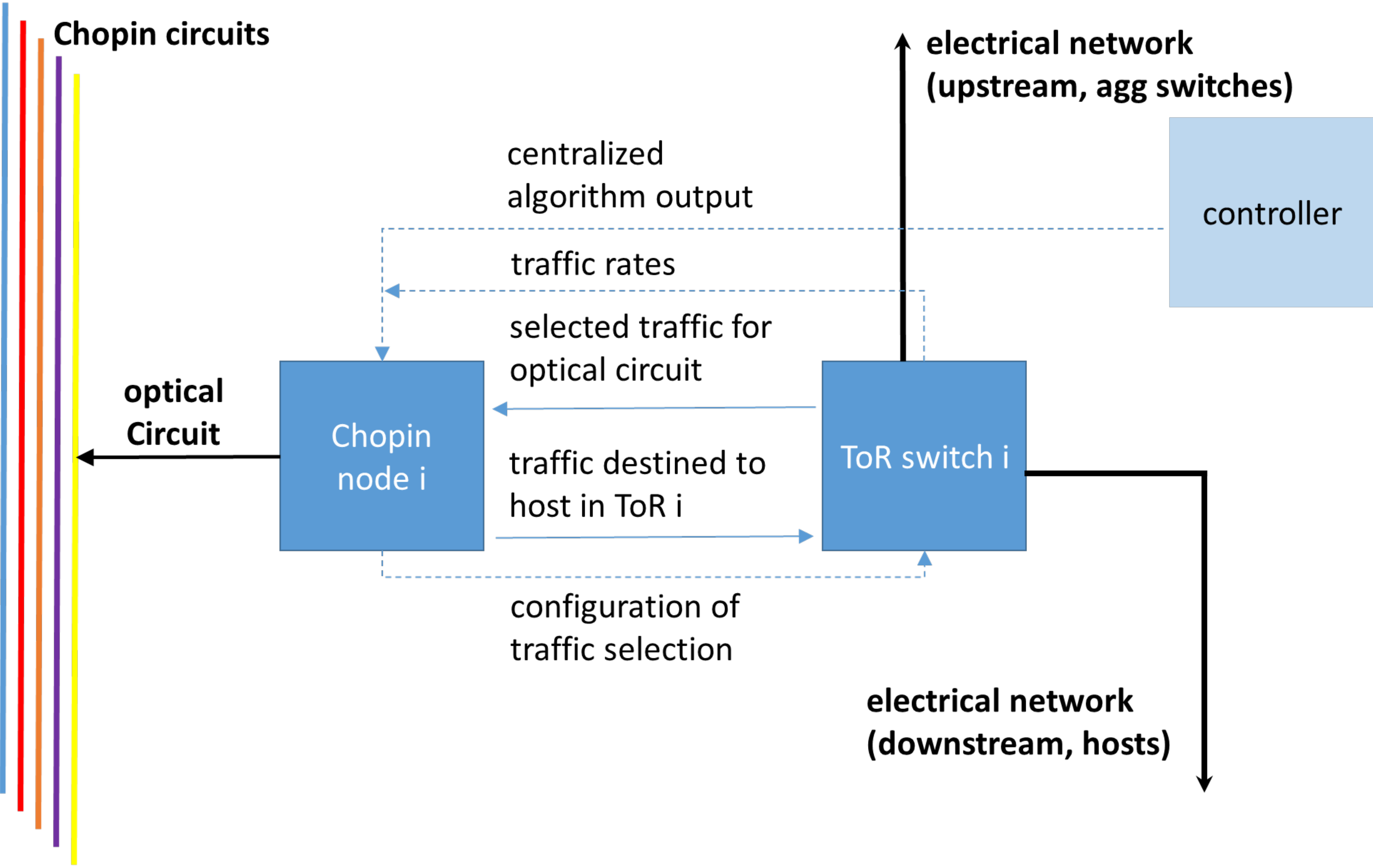}
%	\vspace{-4mm}
	\caption{Chopin's design.}
	\label{fig:distributed_scheduler_fig}
	%\vspace{-5mm}
\end{figure}

Specifically, \systemname's scheduler uses two different control mechanisms, each  carried out in a different location in the datacenter, providing different latency and response times.
		The {\bf centralized scheduler}	is reminiscent of an SDN controller and allows \systemname\ to adapt to global changes (such as traffic rates). 
		This optimization uses traffic measurements across the network and has a (relatively) long response time. Moreover, it may receive additional information (e.g., from applications that have specific repetitive patterns) to make even better decisions. Fig.~\ref{fig:distributed_scheduler_fig} presents the connectivity between the SDN controller to each of the ToR switches and Chopin's nodes. 
The {\bf distributed scheduler} is embedded within the ToR switches and is based only on local measurements. It reacts quickly to local changes in traffic and may tear down connections if they become unmatched and establish new connections for new ``hot'' ToR switch pairs. The tear down and connection establishment are made by updates sent from the ToR switch to its Chopin node, see Fig.~\ref{fig:distributed_scheduler_fig}. 
%
%\begin{comment}
%\end{comment}

\begin{comment}
\begin{wrapfigure}{r}{0.6\textwidth}
\vspace{-7mm}
  \begin{center}
    \includegraphics[width=0.5\textwidth]{figs/local_scheduler_fig1.pdf}
  \end{center}
  \vspace{-5mm}
  \caption{Chopin's design.}
  \label{fig:distributed_scheduler_fig}
\end{wrapfigure}
\end{comment}

%
The centralized scheduler and the distributed scheduler are discussed in details in Section~\ref{sec:centralized} and Section~\ref{sec:distributed} respectively.
\begin{comment}
To enable a centralized scheduler, each of the Chopin nodes is also connected to a logically-centralized controller.  
The controller is also connected to each of the ToR switches, in order to coordinate their interactions (see Fig.~\ref{fig:Chopin_Concept_3d}).
\end{comment}

Moreover, by combining these two schedulers, we can strike an optimized trade-off and realize both fast reactions and global and long-term network optimizations, accounting for demand uncertainty.
In particular, unlike many existing solutions, which consider only one scheduler, Chopin is flexible and performs better than both.

\subsection{The Hybrid\protect\footnote{We note that the term \emph{hybrid} can have a different meaning in some networking contexts, e.g., indicating a combination of the \textsc{LOCAL} model with the node-capacitated clique model~\cite{DBLP:conf/soda/AugustineHKSS20}.} Topology}

Chopin can be used together with any fast switching circuit technology (as in \cite{Mordia,TMS1,huntingMice-TMS1}), and implemented\ignore{using interlacing} within the existing datacenter hardware. 
We distinguish between two entities in the ToR switches: the electric switch itself (for brevity, we will simply refer to this switch as the ToR switch), and the Chopin node which resides in the switch, serving as the entry point to the optical network. This modular Chopin structure enables us to support existing ToR switches, by directing one of its upstream ports to the Chopin nodes. When clear from the context, we use the terms, ToR switch and Chopin node,~interchangeably. 

The optical network can be any non-blocking topology, where the only constraint on establishing a circuit between two ToR switches is the availability of a transceiver in the corresponding Chopin node (namely, its optical degree). 
\ignore{Fig.~\ref{fig:Chopin_Concept_3d} depicts an example network where the ToR switch pairs are inter-connected through Chopin nodes to an optical ring, as well as through an electric packet-switching network.}

Specifically, we assume each Chopin node has an optical degree of $k$ and optical circuits are symmetric. This implies, that at any given time, a Chopin node can send and receive data \emph{from at most $k$ Chopin nodes}. For any given time $t$, we denote by $\textit{dest}_i(t)$ the set of Chopin nodes connected to the Chopin node $i$. We observe that as circuits are symmetric, if $j\in \textit{dest}_i(t)$ then it also holds that $i\in \textit{dest}_j(t)$.

\subsection{Problem Formulation} \label{sec:formulation}

At the heart of Chopin lies the desire to improve network performance and throughput
by avoiding scheduling bottlenecks. 
%To this end, Chopin relies on a centralized and a distributed network optimizer. 
As Chopin is deployed between ToR switches, the scheduler 
is oblivious to intra-rack traffic or delays.

We first need to introduce some preliminaries.
Let $n$ be the number of ToR switches in the network
and assume that time is slotted, where in each time-slot the distributed scheduler can be invoked (e.g., the length of each time-slot is 1 ms). 
%We can refer to the time-interval  $[t,t+\Delta t)$ as epoch $t$. 
Let $X_{i,j}(t)$ be the total amount of traffic sent from rack $i$ to rack $j$ at time-slot $t$. 
Now let $Y_{i,j}(t)$ be an indicator variable to describe whether a pair of ToR switches is connected through a Chopin circuit at time interval $t$:  $Y_{i,j}(t)=1$ if and only if $j\in \textit{dest}_i(t)$ (and 0 otherwise). If $Y_{i,j}(t)=1$ then $Y_{j,i}(t)=1$ as connections through Chopin are symmetric. 

Let $C_t\subseteq S\times S$ be a symmetric relation with all ToR switch pairs that are connected through a Chopin circuit at time interval $t$ (i.e., $(i,j) \in C(t)$ if and only if $j\in \textit{dest}_i(t)$).

%Similarly, we use $Y_{i,j}(t)$ as an indicator variable to describe $C(t)$:  $Y_{i,j}(t)=1$ if and only if $(i,j)\in C(t)$ (and 0 otherwise). 

We aim to maximize optical circuit throughput, a standard objective in such topologies~\cite{RotorNet,c_Through,Helios,REACToR,FireFly,Mordia,Megaswitch,eclipse}, namely
$\sum_t\sum_i\sum_j X_{i,j}(t)\cdot Y_{i,j}(t)$.
This relieves the electrically switched network part and reduces the overall latency. This is done by updating the set~$C_t$, based on local and centralized decisions. 

Note that, as optical circuit capacities are typically very high, we assume that the capacity of an optical circuit is always larger than the total amount of traffic sent between two racks (namely $X_{i,j}(t)$). In case this does not hold, and the two racks are connected through an optical link, one can send traffic through the optical circuit up to its capacity while the remaining traffic is sent through the electrically switched network. 

\subsection{Schedulers and Definitions}

First, a \emph{centralized} scheduler has a global view of the network and, in some cases, even auxiliary information given by the network administrator. This, on one hand, enables the scheduler to perform more informed decisions. But on the other hand, when using a centralized scheduler, it can take much longer to gather, compute, and spread the information across the datacenter. In our model, we assume the centralized scheduler works every $T$ time-slots (which we call the \emph{centralized scheduler epoch}) and uses slightly outdated traffic information: at time $t$, only the measurements $\{ X_{i,j}(t') | t'<t-\Delta, \mbox{for every } i,j\}$  can be used, where $\Delta$ is the \emph{centralized algorithm delay}: the time it takes it to gather all information and make decisions. For example, if the optical degree is $1$ (i.e. $k=1$), the centralized scheduler may use algorithms such as maximum weight matching to optimize the throughput that goes through the optical circuits.  

\ignore{
\begin{figure}[tb]
  \centering
  \includegraphics[width=0.65\linewidth]{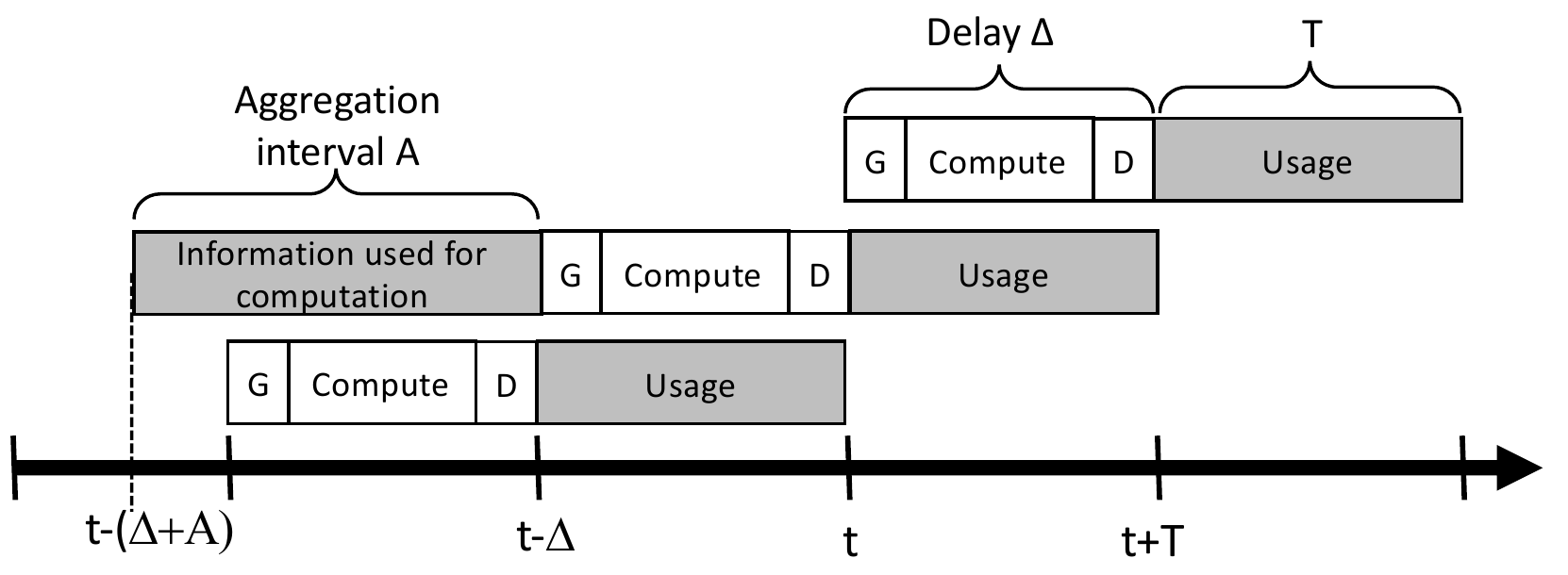}
  \vspace{-3mm}
  \caption{Centralized scheduler timing parameters, where ``G'' stands for the gathering period and ``D'' for the disseminate period. Similar parameters are used  by the \emph{distributed scheduler}, with a delay of~$\delta$. }
 \label{fig:centrelized_parameters}
\vspace{-5mm}
\end{figure} 
}

As $T$ becomes larger, centralized scheduler decisions can deteriorate, as the input on which decisions are based is outdated toward the end of the epoch. Thus, we additionally consider a distributed scheduler that is more fine-grained and runs every time-slot, benefiting from a reduced computation time and avoiding the delays involved in the centralized scheduler; it changes the pairs of connected switches based on local information only and by exchanging messages between ToR switches in two rounds.
%(e.g., similar to iSLIP). 
Specifically, the distributed scheduler of node $i$ at time-slot $t$ may use traffic measurements on its node until the computation starts: 
\vspace{-1.5mm}
\[\{ X_{i,j}(t') | j\neq i, t'<t-\delta \} \cup \{ X_{j,i}(t') | j\neq i, t'<t-\delta \},\] 
where $\delta<\Delta$ is the distributed scheduler delay. In addition, the distributed scheduler is aware of the information sent to it by other nodes throughout the rounds of computation. Importantly, for each pair $(i,j)$ that was optically connected through Chopin at time interval $t-1$ (namely, $Y_{i,j}(t-1)=1$), the distributed scheduler at node $i$ knows what information was used to establish this connection (e.g., what is the rate reported to the centralized scheduler upon its last invocation) and decides whether the information is stale or not. 
Table~\ref{table:notations} summarizes Chopin schedulers' notations.

\begin{table}[h]
\centering
\scriptsize
%\resizebox{\columnwidth}{!}{
\begin{tabular}{p{1.3cm} p{9.5cm}}
%\hline
\specialrule{.1em}{.05em}{.05em} 
Notation & Meaning  \\
\hline
$n$ & The number of ToR switches. \\
$dest_i(t)$ & The set of racks optically connected to rack $i$ at time-slot $t$ \\ 
$X_{i,j}(t)$ & The total amount of traffic sent from rack $i$ to rack $j$ at time slot $t$ \\ 
$Y_{i,j}(t)$ & Indicator variable. $Y_{i,j}(t)=1$ iff $j\in dest_i(t)$ \\
$C_t$ & The set of rack pairs optically connected at time slot $t$ \\
%\hline
$K$ & Optical degree, the number of available circuits per Chopin node.\\ 
%\hline
$\Delta$ &  Centralized scheduler delay\\
%\hline
$\delta$ &  Distributed scheduler delay\\
 %\hline
$\alpha$ &  Chopin threshold for keeping centralized decisions \\
%\hline
$A$ &  Centralized scheduler aggregation interval \\
%\hline
$a$ &  Distributed scheduler aggregation interval \\
%\hline
$T$ & Centralized scheduler epoch\\
%\hline
\specialrule{.1em}{.05em}{.05em} 
\end{tabular}
%}
\caption{Chopin's \ignore{centralized and distributed} schedulers' notations}
\label{table:notations}
%\vspace{-8mm}
\end{table}

\begin{figure}[b]
% \vspace{-3mm}
  \centering
  \includegraphics[width=0.75\linewidth]{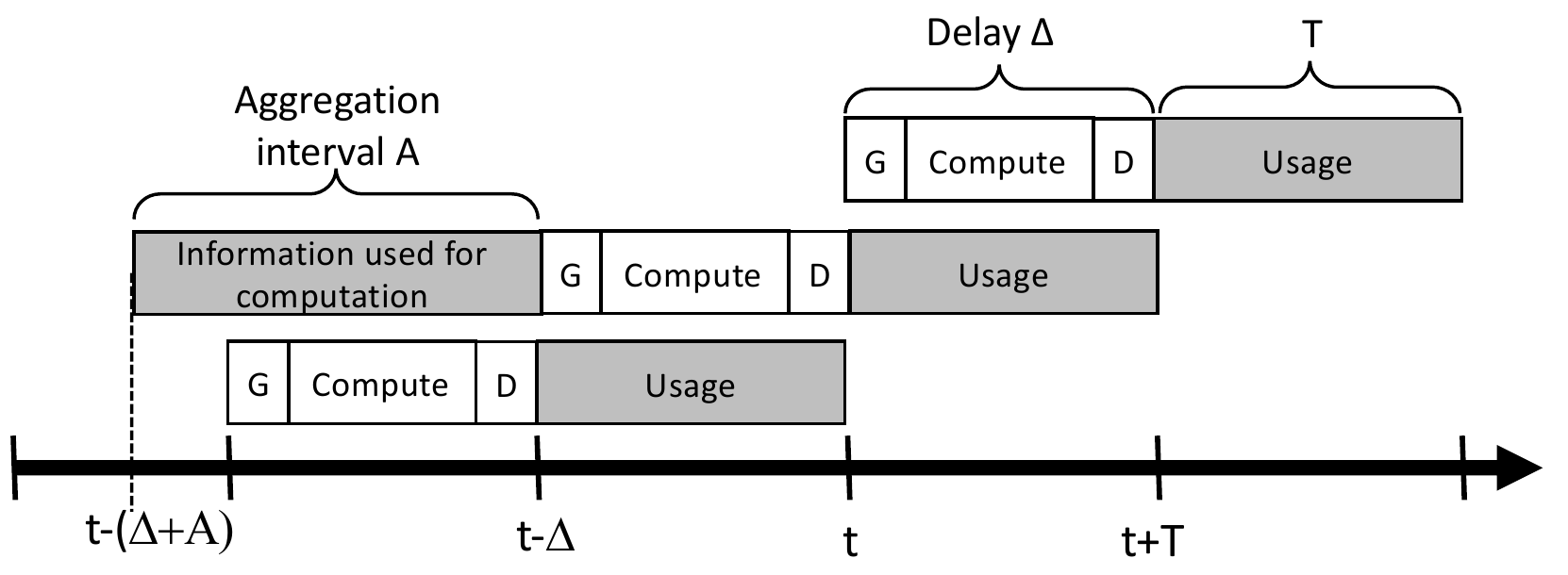}
%  \vspace{-3mm}
  \caption{Centralized scheduler timing parameters, where ``G'' stands for the gathering period and ``D'' for the disseminate period. Similar parameters are used  by the \emph{distributed scheduler}, with delay of~$\delta$. }
 \label{fig:centrelized_parameters}
%\vspace{-5mm}
\end{figure}

\section{Chopin's Centralized Scheduler} \label{sec:centralized}

The centralized scheduler is implemented on top of a centralized SDN controller, which is (logically) connected to each of the ToR switches and the Chopin nodes. %through optical circuits. 
Upon a request from the centralized scheduler, the controller collects traffic measurements across the network (namely, counters at ToR switches, current status of Chopin nodes).
Based on these measurements, it computes the next optical circuit allocation.

Recall that the \emph{delay $\Delta$} is the time it takes to send all the information to the controller, run the centralized algorithm, and send the decisions back to the nodes.
The centralized algorithm works in epochs of length $T$, where decisions arrive at the nodes at the beginning of each epoch.
Assume an epoch starts at time $t$.
Then, these decisions will be used by nodes until time $t+T$ (or until altered locally by the distributed scheduler).
Furthermore, these decisions are based on information gathered in the interval $[0,t-\Delta]$.
However, if traffic changes quickly (DC traffic is often bursty~\cite{sigmetrics20complexity}), this information may be outdated quickly. This motivates us to define an \emph{aggregation interval $A$} for the centralized algorithm, considering only the interval $[t-(\Delta+A), t-\Delta]$, see Fig.~\ref{fig:centrelized_parameters}.

Notice that the delay $\Delta$ is an important factor for the performance of Chopin. 
The delay describes the response time of the central scheduler and consists of several steps:
contacting tens to hundreds of nodes~\cite{FlywaysCongestedDC,task_datacenter}, (2) receiving thousands of flow entry statistics, estimating optical circuit utilization, contacting all nodes again, and updating all rules with new parameters if needed.

Considering common SDN controllers' capability to handle a few thousands of messages per second \cite{controller-perf}, we estimate the delay to be in the order of hundreds of milliseconds in most configurations \cite{NeedToKonwControlPlaneConfigTime2} \cite{MeasuringConfigTime3}.
Furthermore, the computation time of our algorithms can be in the order of tens of milliseconds for hundreds of ToR switches (e.g., when running maximum weight matching--like algorithms, as reported in \cite{Helios}). %But all in all, the reconfiguration cycle still suffers from long~delays.
%
%This factor delays are 
%summarized in 
%see~Table~\ref{table:ControlPlaneLatencies}.
%
 Due to these delays, fast changes in the network (occurring within a few milliseconds \cite{Facebook_2015}),  may not be detected by the centralized scheduler in a timely manner. Also, the reconfiguration time (approximately $11$ $\mu$sec \cite{projectToR} \cite{TMS}) is likely negligible compared to a centralized reconfiguration cycle. These observations motivate usage of another scheduling layer, to adapt to traffic in an online~manner.

%\subsection{Matching based algorithms}
Our high-level goal is to maximize the overall throughput over the optical network. First recall that  allocations are constrained by the optical ToR switch degree~$k$: each ToR switch can be optically connected to at most $k$ other ToR switches, Accordingly, our centralized algorithm essentially needs to solve a weighted b-matching problem, with $b=k$. Specifically, we consider an undirected graph whose nodes are the ToR switches and the weight of each edge $(i,j)$ is the total traffic between $i$ and $j$ in the relevant interval: 
\vspace{-4mm}
\[ 
w_{ij} = \sum_{t'=t-(\Delta+A)}^{t-\Delta} X_{i,j}(t')+X_{j,i}(t').
\]
\vspace{-4mm}

While b-matching algorithms are strongly polynomial~\cite{DBLP:journals/ipl/Anstee87}, their running time can still be prohibitively high in practice~\cite{DBLP:journals/jea/Muller-HannemannS00,DBLP:journals/siamdm/LetchfordRT08,DBLP:journals/talg/Gabow18}. This can lead to high delays $\Delta$ and in turn, to a significantly reduced 
overall performance of the system.
Thus, we propose to approximate the problem: we compute a maximum weight matching (using Edmond's MWM algorithm \cite{Edmonds_imp}), subtract the weights of the matching's edges from the graph, and run maximum weight matching again with the new, smaller weights. As in \cite{OSA}, this process is repeated $k$ times, resulting in $k$ matchings. Hence each node is connected to at most $k$ other nodes, as required. We refer to Khan et al.~\cite{DBLP:journals/siamsc/KhanPPSSMHD16} for a further discussion on efficiently approximating b-matchings.

%A similar greedy approximation, resulting in significant computation time reduction, was used also in~\cite{OSA}.

We further reduce computation time by considering only the top-$m$ live flows per ToR switch (instead of all possible pairs between the nodes). Due to the sparse nature of datacenter traffic matrices, even small values of $m$ provide a highly accurate approximation: there is almost no performance degradation compared to a full-fledged MWM (\S5). %(see our experiments in Section~\ref{sec:evaluation}).
Note that the top-$m$ flows per switch can be efficiently calculated in each switch since there are only $n$ possible flows and maintaining $n$ counters at line rate is supported by switches.

\begin{figure}[t]
%\vspace{-8mm}
\begin{minipage}[b]{0.48\textwidth}
\centering
\includegraphics[width=\textwidth]{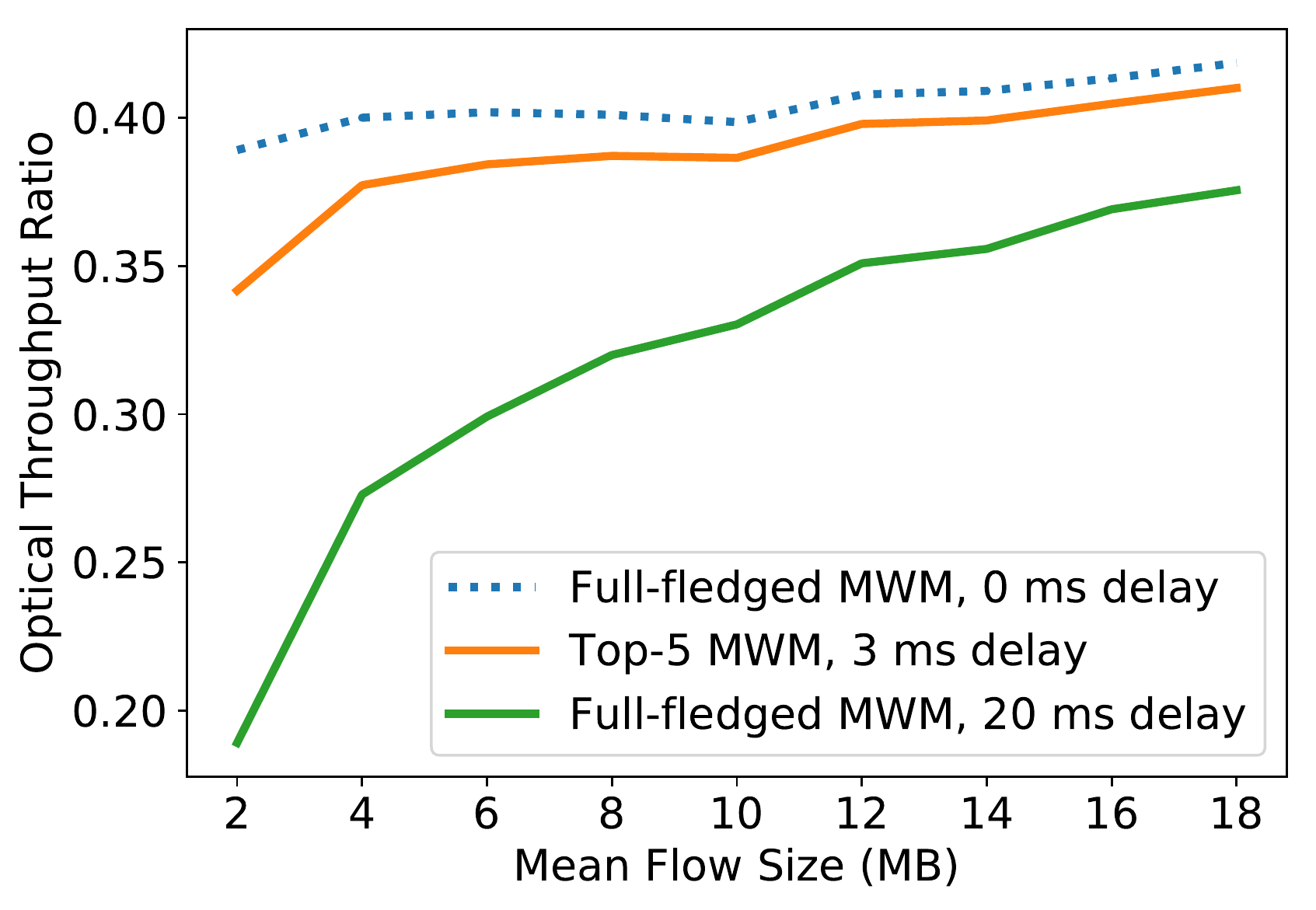}
\caption{Comparison between a centralized scheduler which operates every $3$ ms and computes the MWM of the top-$5$ live flows, to a centralized scheduler which operates every $20$ ms and computes full-fledged MWM. 
  For brevity only pFabric results are shown.}
 \label{fig:centrelized_schedulers_row1}
\end{minipage}
\hfill
\begin{minipage}[b]{0.48\textwidth}
\centering
\includegraphics[width=\textwidth]{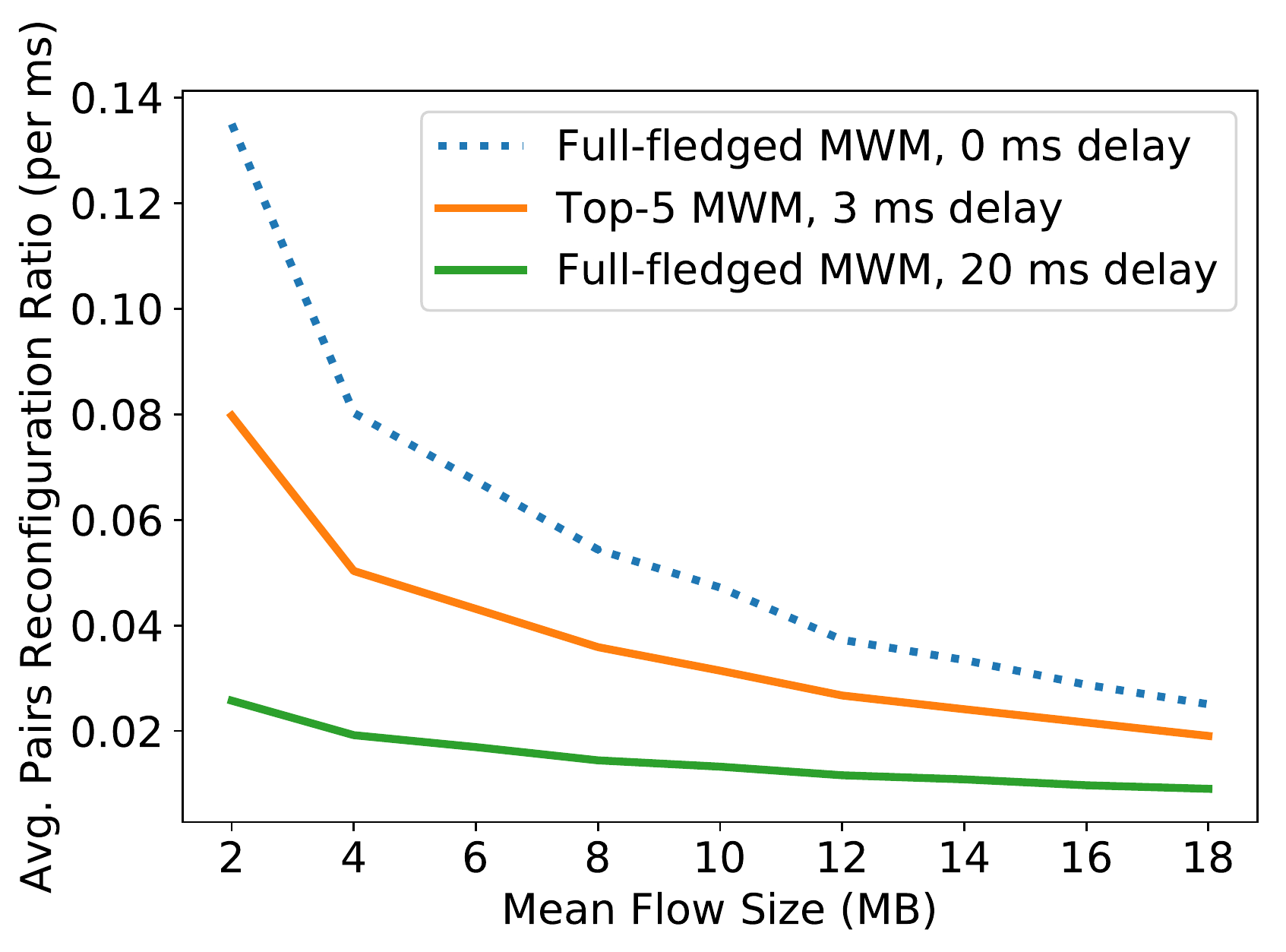}
 \caption{Average reconfiguration ratio per $1$ ms of three scenarios: centralized scheduler which %\neta
 computes MWM over top-$5$ live flows every $3$ ms, a centralized scheduler which operates every $20$ ms and computes full-pledged MWM. 
  For brevity only results for the pFabric traffic pattern are shown.}
 \label{fig:mega_comb_heatmap_centralized_3}
\end{minipage}
\end{figure}

Moreover, focusing only on a constant number of top flows per node enables Chopin to scale with an increasing number of nodes. It also decreases both the MWM computation and the network reconfiguration times, allowing more frequent centralized scheduler reconfigurations. As the frequency of centralized scheduler invocations significantly affects the performance, by considering only $m$ flow, we can improve the scheduler's performance. 
For example, when $m=5$ and the number of ToR switches is $80$, the time it takes to compute MWM based on top-$5$ live flows per ToR is $3$ ms, while full-fledged MWM takes at least $20$ ms.  Fig.~\ref{fig:centrelized_schedulers_row1} compares the performance of both algorithms under the pFabric traffic pattern we have described (similar results hold for other traffic patterns as well) and shows that having more frequent reconfigurations is more significant than having slightly better matchings. 
Our centralized scheduler, based on top-$5$ live flows with reconfiguration every $3$ ms, achieves almost the same results as an idealized online optimal algorithm, that computes full-fledged MWM every $1$ ms. 
Finally, we observe that the optical throughput ratio improves as the mean flow size increases (and the gap between the algorithms shrinks), since longer flows imply that flow information is still relevant even after a long time when computations are infrequent.

In order to explain the throughput differences in Fig.~\ref{fig:centrelized_schedulers_row1}, we analyzed the number of reconfigurations in each scenario. We consider the average number of reconfigured pairs in  each run, out of the total number of pairs ($n/2$). Fig.~\ref{fig:mega_comb_heatmap_centralized_3} presents the reconfiguration average ratio per $1$ ms, as a function of the mean flow size. As expected, as the mean flow size increases (and the flows are longer), the number of reconfigurations decreases. Moreover, it shows that  the number of reconfigurations is decreasing rapidly when the epoch time is $20$ ms. This can be attributed to the fact that short lived connections have less impact on the $20ms$ long measurements and are less likely to be matched.

\begin{comment}
    
\begin{figure}[tb]
\vspace{-8mm}
  \centering
  	\includegraphics[width=0.33\textwidth]{figs/mega_comb_heatmap_centralized_3-20_pFabric_1row.pdf}
  	%\vspace{-5mm}
  \caption{Comparison between a centralized scheduler which operates every $3$ ms and computes the MWM of the top-$5$ live flows, to a centralized scheduler which operates every $20$ ms and computes full-fledged MWM. 
  For brevity only pFabric results are shown.}
  
 \label{fig:centrelized_schedulers_row1}
%\vspace{-4mm}
\end{figure} 

\begin{figure}[tb]
  \centering
  	\includegraphics[width=0.33\textwidth]{figs/mega_comb_heatmap_centralized_3-20_reconf_pFabric_1row.pdf}
  	%\vspace{-5mm}
  \caption{Average reconfiguration ratio per $1$ ms of three scenarios: centralized scheduler which operates every $3$ ms and computes the MWM of the top-$5$ live flows, to a centralized scheduler which operates every $20$ ms and computes full-pledged MWM. 
  For brevity only results for the pFabric traffic pattern are shown.}
 \label{fig:mega_comb_heatmap_centralized_3}
%\vspace{-3mm}
\end{figure} 
\end{comment}

\section{Chopin's Distributed Scheduler} 
\label{sec:distributed}
%\vspace{-1mm}
The distributed scheduler is a distributed control algorithm, embedded inside each Chopin node.
Each ToR switch is connected to a single Chopin node and sends flows to the latter (e.g., by connecting one of its ports to the Chopin node). 
The traffic that is sent through this port is configured either by our centralized algorithm (as described in Section~\ref{sec:centralized}) or by the distributed scheduler that runs on the Chopin node. 

The Chopin node is responsible of sending traffic destined for the ToR switch from one of the optical circuits. 
%As each ToR switch is connected to a single Chopin node, overheads per node are small.  
Each ToR switch in turn is connected to a single Chopin node.
We refer to the illustration in Fig.~\ref{fig:distributed_scheduler_fig} for an overview.
Supplemental pseudo-code of Chopin's distributed algorithm appears in Algorithm~\ref{alg:distributed_alg} in the Appendix.

At the beginning of each centralized scheduler epoch, every Chopin node keeps track of the traffic rate according to which its circuit was selected. Namely, for an epoch that starts at time $t$, if a circuit was established between Node $i$ and Node $j$, both Node $i$ and Node $j$ compute and store: 
%\vspace{-2.5mm}
\[ R_{i,j}(t) = \frac{1}{A}  \sum_{t'=t-(\Delta+A)}^{t-\Delta} X_{i,j}(t')+X_{j,i}(t'). \]
%\vspace{-2mm}

The nodes use these rates to determine if traffic demands stay steady during the epoch. Specifically, we define a \emph{threshold} $\alpha \geq 0$, and compute the rate in each time-slot $\hat{t}\in[t,t+T]$ based only on local information available at the nodes: 
%\vspace{-1mm}
\[ r_{i,j}(\hat{t}) = \frac{1}{a}  \sum_{t'=\hat{t}-(\delta+a)}^{\hat{t}-\delta} X_{i,j}(t')+X_{j,i}(t'), \]
%\vspace{-3mm}

\noindent where $\delta$ is the distributed scheduler delay, and $a$ is its aggregation interval.
Only if $r_{i,j}(\hat{t}) > \alpha \cdot R_{i,j}(t)$, then the circuit is marked as matched 
%(Algorithm~\ref{alg:distributed_alg}, Lines 2--5)
and the algorithm keeps it connected through this epoch. Otherwise, it strives to replace it with a better connection, as described next. 
We observe that for $\alpha=0$, all existing connections are kept matched (namely, Chopin just runs the centralized scheduler, and may improve only when its computed b-matching changes).
As the threshold increases, it enables the distributed scheduler to tear down almost every centrally-computed connection and to create new ones, based on the current ToR switch traffic. 
The distributed scheduler itself tries to establish as many circuits as possible to increase the overall traffic through the optical circuit connections.
%
%First, each Chopin node $i$, consider his connection matched by the controller. Out of these connections, every connection $j$ which is matched by the controller with bi-directional traffic which is above $\alpha \cdot R_{i,j}(t)$ is marked as \texttt{matched peer}. 
%Then, the \texttt{free links} are calculated as the number of allowed connection per Chopin node ($k$) minus the number of matched peers (line $19$).
%
In a nutshell, each Chopin Node $i$ sends requests to a predetermined number of other nodes needs (this number is denoted by variable max\_reqs), for which it observes the most bi-directional traffic. These nodes, denoted by req\_nodes,
%(Line $7$), 
do not include those kept matched to Node $i$; moreover, max\_reqs $
> k$ to allow utilizing all the circuits connected to Node $i$. 
After a Chopin node $i$ sends its requests, it waits to receive requests from other nodes. We distinguish between: \ignore{two cases: }
\begin{enumerate}
    \item Request from a node $j$ that is in the req\_nodes set: This means that both nodes $i$ and $j$ consider the traffic between them in their top max\_reqs links. This makes Node $j$ a candidate for a match with Node $i$. 
    \item Request from a node $j$ that is not in the req\_nodes set: This means that while Node $j$ considers Node $i$ in its top max\_reqs links, Node $i$ has max\_reqs other links with larger traffic. This request should be denied. 
\end{enumerate}

We wait until all requests are received at Node $i$: This is indicated by a time-out event, that can be set, for example to half the aggregation interval $a$ (requests are timestamped, so requests that arrive after the time-out will simply be ignored.). After all requests are received, there will be at most max\_reqs candidates for matching.
%(Line $13$). 
However, the number of free circuits (the optical degree minus the number of matched circuit) may be smaller. Therefore, we choose only the top ones so as not to exceed the number of available links.
%(Line $15$). 
We thus send a \texttt{grant} message to all of them 
%(Line $18$) 
and \texttt{deny} messages to others.
%~(Line~$17$). 
 
In the last phase of our algorithm, each node waits until all its requests are either granted or denied. It then connects with all nodes that \emph{(i) it has granted, and (ii) a grant message was received from them}.
%(Line $30$). 
It disconnects all other links, except those made by the centralized algorithm and above the threshold. % as determined in Lines $2$--$5$.  
Note that rate measurements used in an epoch are performed in parallel with the decision making of the previous epoch.

\begin{comment}
\begin{figure}[tbp]
	\centering
	\includegraphics[width=0.45\textwidth]{figs/local_scheduler_fig1.pdf}
	\caption{Chopin's distributed scheduler scheme}
	\label{fig:distributed_scheduler_fig}
\end{figure}   
\end{comment}

\section{Evaluation} \label{sec:evaluation}
Chopin aims to maximize the circuit throughput (online), without compromising the datacenter latency, by combining centralized and distributed schedulers.
Therefore, in our evaluation, we focus on each of these schedulers' parameters as well as on their contribution to the overall DC performance.
The performance is evaluated on several parameters, including the centralized scheduler epoch $T$, aggregation intervals $A$ (for the centralized scheduler) and $a$ (for the distributed scheduler), as well as the corresponding delays $\Delta$ and $\delta$. 
%    \item \emph{Compute epoch} $T$: the time between centr computations.
%    \item \emph{Aggregation interval} : the time interval considered by a MWM computation.
%    \item \emph{Window}: the time interval considered by each distributed computation step.
%    \item \emph{Delays}: difference between considered intervals and their usage (aging). The delay applies to % aggregation interval and window.
%\end{enumerate}
%These parameters are presented in \todo{Fig.~\ref{fig:main_params}}.
%

%We discuss scheduler implementation details in \S\ref{subsec:sd}, %and the topology and traffic assumptions of our evaluation in %\S\ref{sec:Implementation}.
%%, and the centralized computational speed-up of Chopin in %\S\ref{sec:Approximations}. 
%We then investigate Chopin's centralized-distributed trade-off in %\S\ref{sec:Trade-off} and perform a sensitivity analysis in %\S\ref{subsec:sensa}.

\subsection{Methodology}\label{sec:Implementation}

\noindent \textbf{Topology.} 
We have analyzed Chopin's performance through synthetic simulations, for which we generate traffic according to known datacenter traffic patterns~\cite{DCIntheWild,Facebook_2015,PoissonTrafficMatrix}. 
We used \emph{NetworkX} for topology creation, as well as for 
%weighted maximal 
 matching computations. % and variants. 
Our simulation code is available at \cite{chopin_code}.

\ignore{
\subsection{Chopin's study Case}
As mentioned above, Chopin fits to any fast reconfiguration technology that enables to connect a circuit between each of the ToR switches in the datacenter. Constructing these circuits requires the availability of devices that can perform fast wavelength selective switching, and electrical wavelength tuning (WSS and EWT, respectively). For a case study for our implementation, we refer to a state of-the-art reconfiguration technology called Dynamic Electroholographic Circuit Switching (DECS). This technology is based on  Electroholography (EH),  implementing WSS and EWT by devices that can be incorporated in Integrated Photonic Circuits (IPC) in which they are interconnected by a mesh of waveguides and operate in unison. This way, DECS enable fast, simultaneous communication between $7,500$ ToR switch pairs in a unicast mode, (approximately $20\%$ higher connectivity than  recent works \cite{Megaswitch}), with $5$ ns reconfiguration time.  

DECS's fibers are connected to all the racks, each with a bandwidth
of $10$ Gbps and a transmission delay of $5$ $\mu s$. }

Specifically, we have considered  real-world datacenter topologies (3-tier) with $8$ and $16$ aggregation switches, $80$ racks and $160$ racks, respectively, where each rack contains $10$ hosts (i.e., up to $1,600$ hosts in the network).
In addition to the electrical network, we assume a non-blocking optical circuit switch, which connects to each Chopin node $k$ times at $10$ Gbps, we vary the value of $k$ between $1$,$2$, and $4$.

Real datacenter's data plane parameters were used. The link capacities are $1$ Gbps between servers and ToR switches, $10$ Gbps between ToRs and aggregation-level switches as in \cite{DCIntheWild}, and $40$ Gbps between the aggregation-level switches and cores, as in \cite{FlywaysCongestedDC}. 
\ignore{The specification of the switches is inspired by 
Dell Switches \cite{DellReport}.} 
The reconfiguration time is approximately $11 \mu sec$~\cite{projectToR,TMS}, as discussed in Section~\ref{sec:centralized}.
As the host's traffic contains hundreds of Mbps on average at all times \cite{PoissonTrafficMatrix}, we analyzed average host demand levels of $200$~Mbps.

Chopin's evaluation focuses on increasing the optical throughput. 
Optimizing Chopin's optical throughput adds some approximation to it, in three aspects: (i) partial maximal weight matching computation, (ii) higher optical degree of Chopin nodes, and (iii) approximated maximal weight matching for higher optical degrees, as discussed next.  

Adding several optical routes per Chopin node improves its optical throughput by using higher connectivity between Chopin nodes. This can be achieved, e.g., by a wavelength-selective switch (WSS) module at each ToR switch, which is a customized $1\times 4$-port Nistica full-fledged $100$ WSS module (as suggested in \cite{MordiaTimeAnalysis}). This implementation enables each Chopin node to connect other Chopin nodes by up to $4$ optical links.

This becomes less attractive for a larger number of channels (namely, greater than $4$)  because of the additional noise (e.g., the multiplexer enables additive noise funneling from each of the sources into the reconfigurable optical add/drop multiplexer ROADM network) \cite{4_connections}. Furthermore, recent studies show that using $1\times 8$ ports increases the system costs by a factor of $10$ compared to $1 \times 4$ ports \cite{4_port_8_port}.
Thus, for cost-effective systems, where several optical switches are recommended, we analyze Chopin's performance where each Chopin node has up to $4$ connections to other Chopin nodes (i.e., ``optical degree~$k=4$''). 

\vspace{1mm}

%\subsection{Traffic Patterns}
\noindent \textbf{Traffic patterns.} 
We generate the traffic flow based on previous studies of traffic characteristics of datacenter networks~\cite{EfficientTrafficMatrix,DCIntheWild,InterRackPattern}. 
Flows are TCP \cite{empirical-traffic-gen} with Poisson flow arrival times~\cite{PoissonTrafficMatrix}, whose size distribution 
follows one of three well-known flow size distributions: (i) HULL \cite{HULL}; (ii) pFabric \cite{DCTCP,pFabric,pFabric2}; and (iii) VL2 \cite{VL2}.\ignore{, as described in Fig.~\ref{fig:three_distributions}.}

The distribution of flow arrival time to the ToR switches is modeled as a Poisson process, where the servers use the network heavily, constantly transmitting and receiving several hundreds of Mbps data on average all the time \cite{PoissonTrafficMatrix}. Such a traffic pattern matches the common inflow rate in today's datacenters serving a variety of applications, such as video and job-task managers \cite{Netflix_requirements,MapReduce_traffic_volume}. 

The dispersion pattern in the simulation was based on the observation
that traffic is either rack-distributed or destined for one  $\approx 1\%{-}10\%$ 
 of the hosts, spread across most of the source's cluster (tens of racks) \cite{InterRackPattern,Facebook_2015}. The inter-rack demand per host was set to approximately $150$ Mbps, based on \cite{PoissonTrafficMatrix}.

Chopin's optical circuits throughput is analyzed w.r.t.: 
\begin{itemize}
    \item Different flow traffic distributions (HULL\ignore{~\cite{HULL}}, VL$2$\ignore{~\cite{VL2}} and pFabric\ignore{~\cite{pFabric}}).
    \item Different scheduler policies: all-distributed/-centralized, and in-between (varying  threshold $\alpha$ levels). %and the spectrum between them (based off varying threshold $\alpha$ levels).
\end{itemize}

We found that each  flow size distribution  has special properties, with respect to flow length and flow size variance. \ignore{, as described in Fig.~\ref{fig:three_distributions}.} These properties have a significant influence on the performance of both the centralized and the distributed scheduler:

\ignore{
\begin{figure}[t]
%\vspace{-3mm}
 \squeezeup
	\centering
	\includegraphics[width=0.3\textwidth]{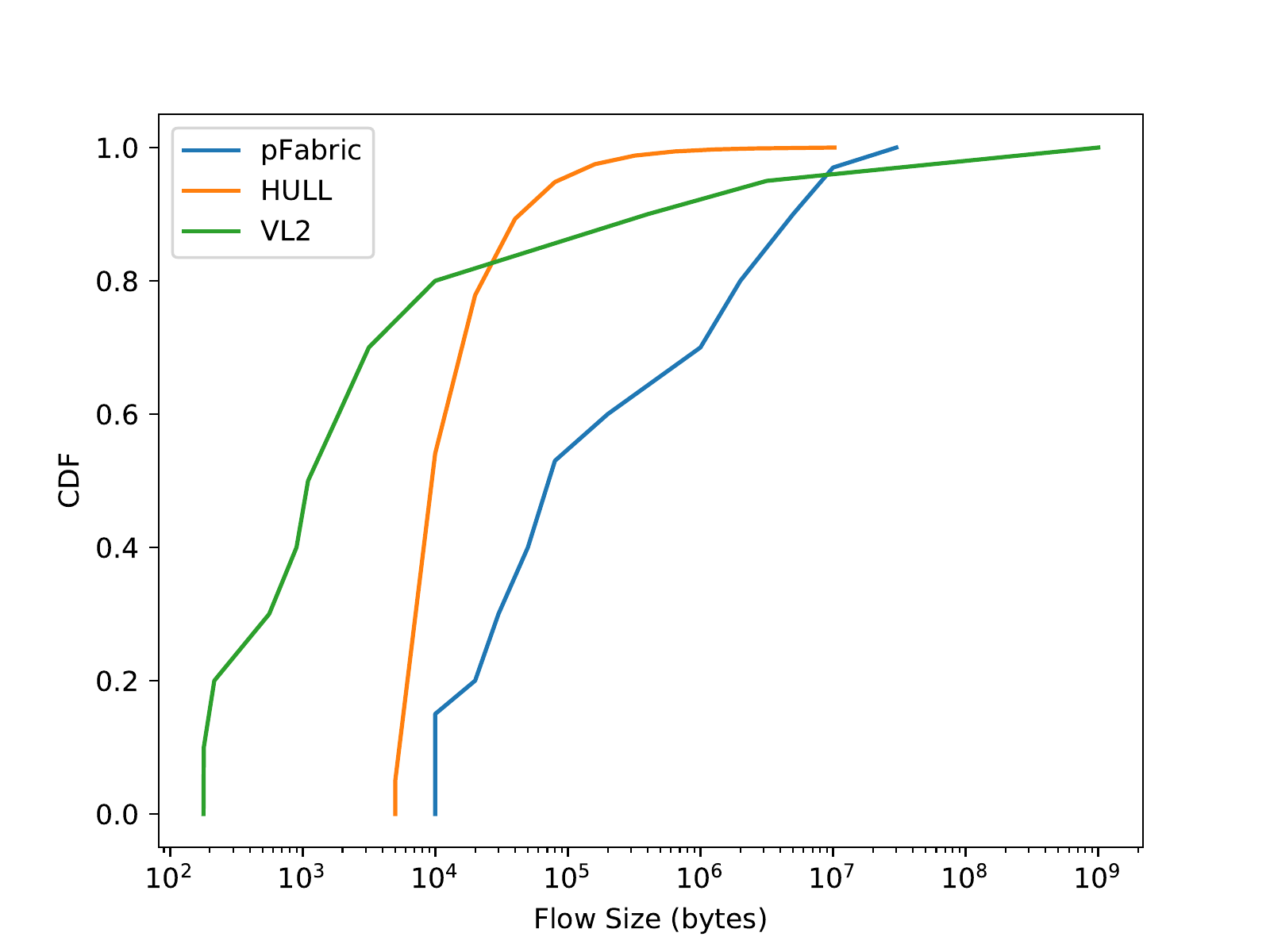}
	%\vspace{-4mm}
	\caption{Flow size distributions}
	\label{fig:three_distributions}
	 %\squeezeup
	 %\vspace{-6mm}
\end{figure}
}

\begin{table}[t]
\centering
\scriptsize
%\vspace{-8mm}
%\resizebox{\columnwidth}{!}{
\begin{tabular}{l  l  l  l} 
\specialrule{.1em}{.05em}{.05em} 
%\hline
Distribution & HULL & pFabric & VL$2$  \\
\hline
%\hline
Mean & Low ($100$ KB) & Medium ($1.7 MB$) & High ($12$ MB)\\ 
%\hline
Variance &  Low & Medium & High \\
%\hline
Centralized perf.
 & $0.42$ & $0.7$  & $0.77$ \\ 
%\hline
Distributed perf.
 & $0.49$ & $0.75$ & $0.77$ \\
%\hline
Chopin
 & $0.5$ & $0.76$ & $0.78$ \\
%\hline
\specialrule{.1em}{.05em}{.05em} 
\end{tabular}
%}
%\vspace{2mm}
\caption{Throughput ratio for optical degree $4$}
\label{table:4_connections_per_tor}
%\vspace{-9mm}
\end{table}

\ignore{
\begin{figure}[t]
	\centering
	\includegraphics[width=0.45\textwidth]{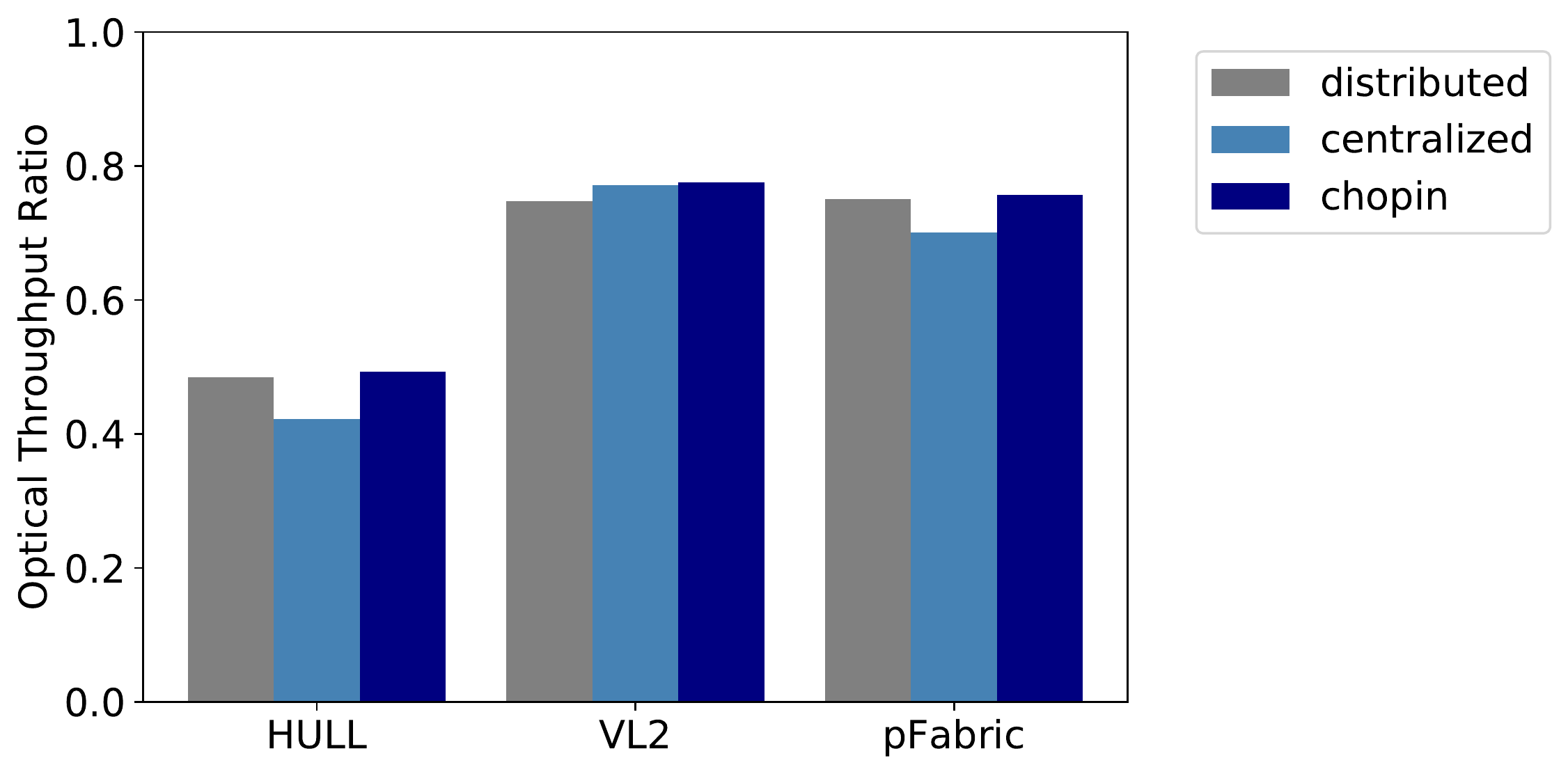}
	\caption{Chopin, all-centralized, and all-distributed scheduler performances for different flow patterns (optical degree $4$).}
	\label{fig:improvemet}
\end{figure}}

\begin{itemize}
    \item The \emph{HULL} flow distribution is a Pareto distribution where almost all flows are mice\footnote{We define mice and elephant flows based on the distinction made by~\cite{DCIntheWild}.} ($<10$KB). Moreover, flow variance is low. Therefore, the centralized scheduler throughput is low, as there are not many elephant flows, and the differences between the flow carried by the optical links is small compared to the others.
    \item The \emph{VL2} distribution creates many elephant flows, with high variance.
	Therefore,  the centralized scheduler can optimize the traffic and
	the distributed scheduler can make decisions which improves the  throughput through the optical circuits.
	\item The \emph{pFabric} distribution includes some elephant flows (but medium mean). With medium variance the distributed scheduler operates as for VL$2$, but centralized is slightly less effective, due to shorter flows.
\end{itemize}

The properties of these traffic patterns and their impact on the scheduler performance are described in Table ~\ref{table:4_connections_per_tor}, for an optical degree of $4$.
Notice that Hull, as a Pareto distribution with $\alpha=1.05$, mean=$100 KB$ \cite{HULL} has unbounded variance. pFabric has mean value of approximately $1.7$ MB \cite{pFabric2} and variance of $3.9$MB,  and VL$2$ \cite{VL2} has mean value of $12$ MB with variance of $85$MB.

\begin{figure*}
%\vspace{-4mm}
\centering
\begin{subfigure}{0.31\linewidth}
  \centering
  \includegraphics[trim=0 0 1.0in 0, clip, width=\linewidth]{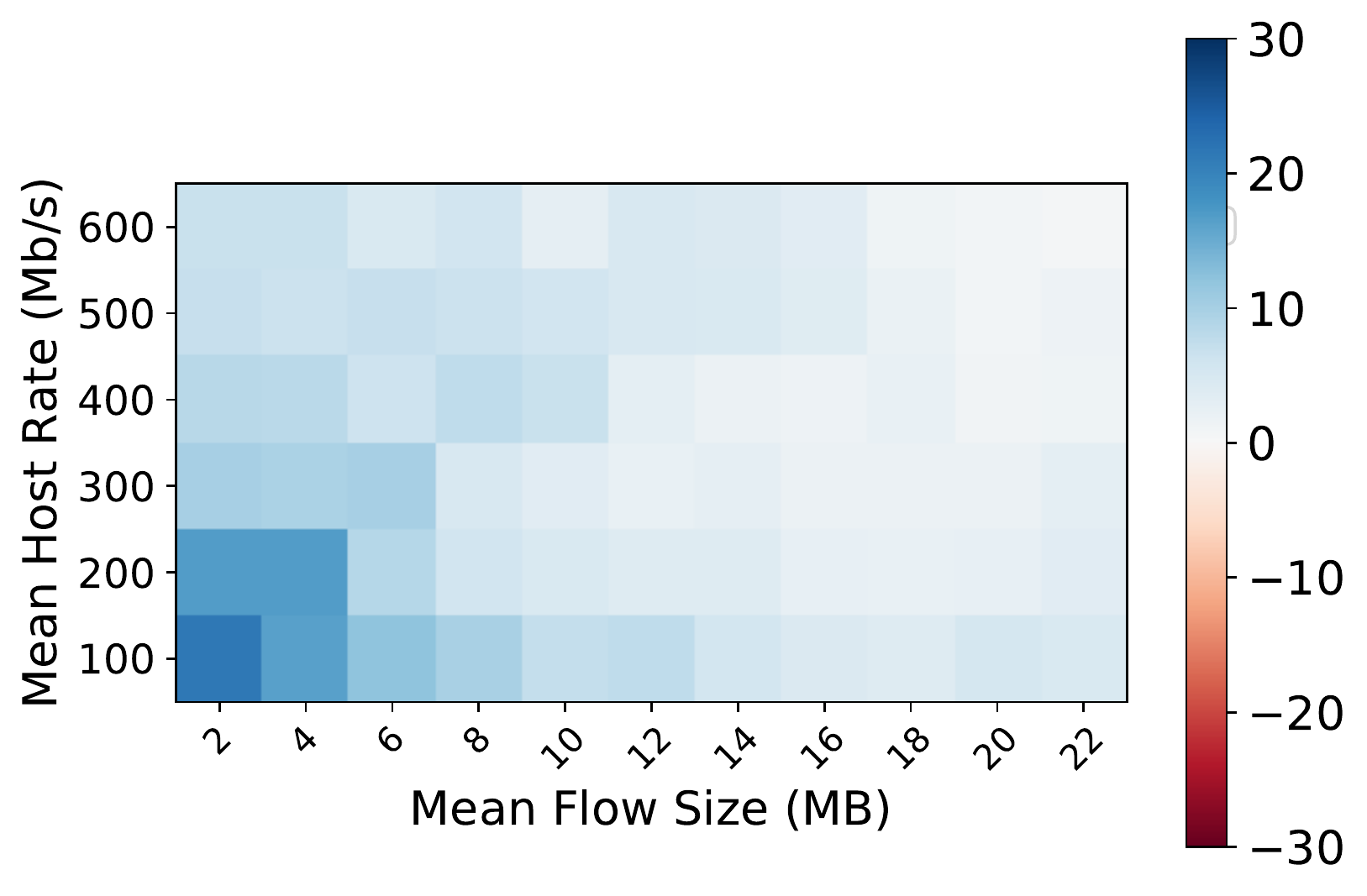}
  %\vspace{-7mm}
  \caption{HULL}
  \label{fig:mega_comb_heatmap_deg4_chopin-centralized_rel_HULL}
  \quad
\end{subfigure}%
\begin{subfigure}{0.31\linewidth}
  \centering
  \includegraphics[trim=0 0 1.0in 0, clip, width=\linewidth]{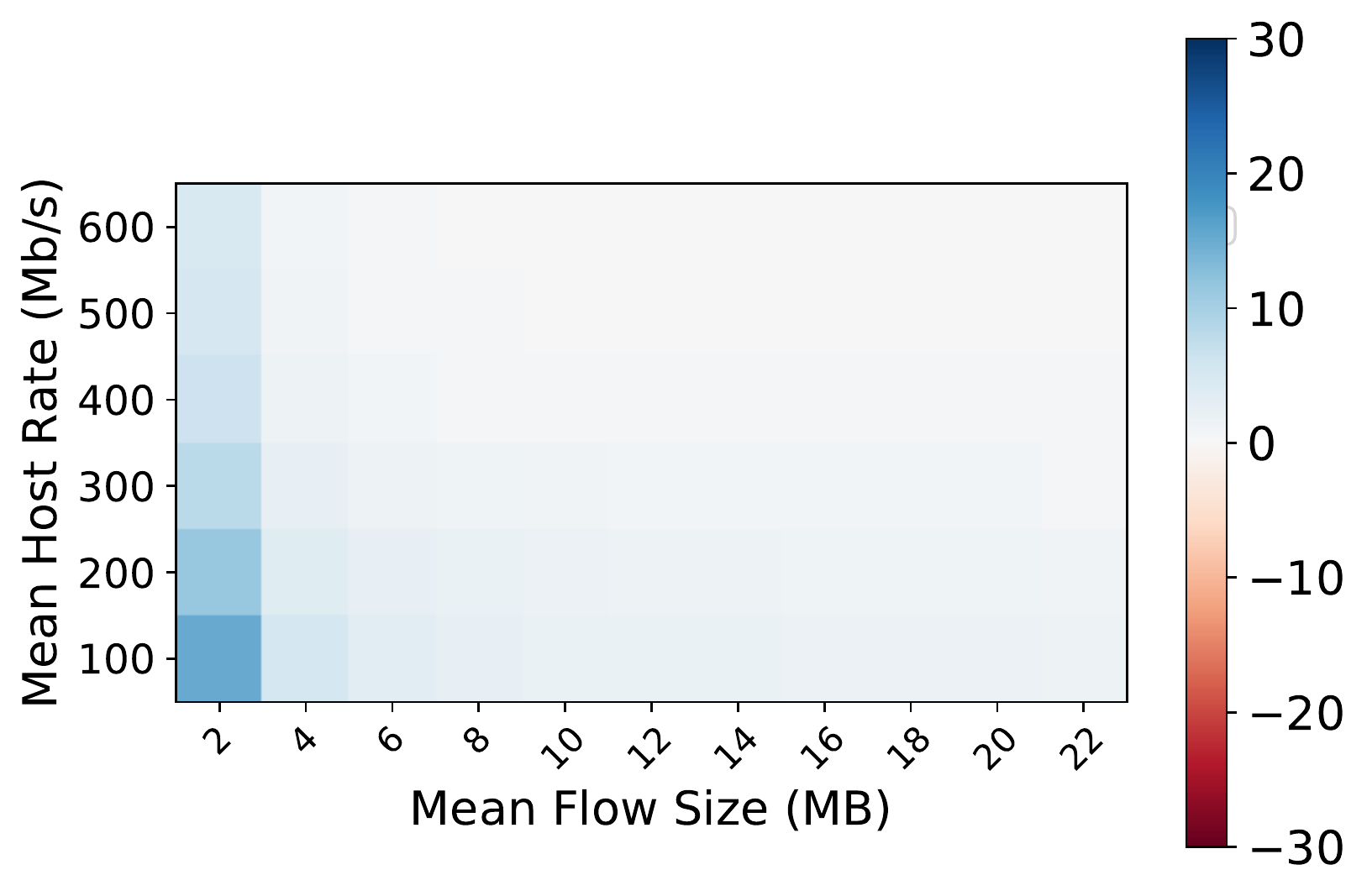}
  %\vspace{-7mm}
  \caption{pFabric}
  \label{fig:mega_comb_heatmap_deg4_chopin-centralized_rel_pFabric}
  \quad
\end{subfigure}%
\begin{subfigure}{0.31\linewidth}
  \centering
  \includegraphics[trim=0 0 1.0in 0, clip, width=\linewidth]{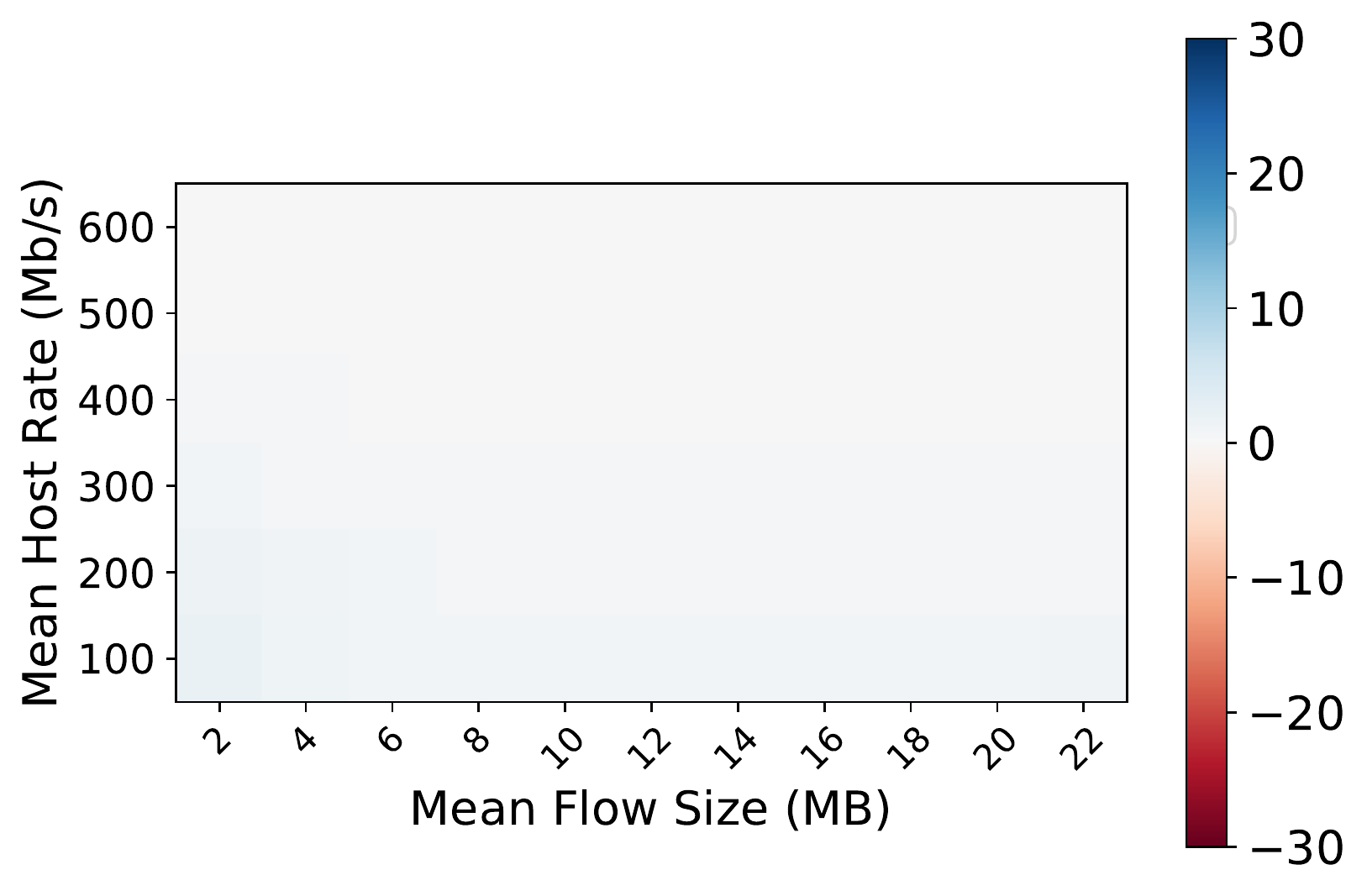}
 % \vspace{-7mm}
  \caption{VL2}
  \label{fig:mega_comb_heatmap_deg4_chopin-centralized_rel_VL2}
  \quad
\end{subfigure}%
\begin{subfigure}{0.05\linewidth}
  \centering
 \includegraphics[%trim=0 0.5in 0 0, clip,
  scale=0.35]{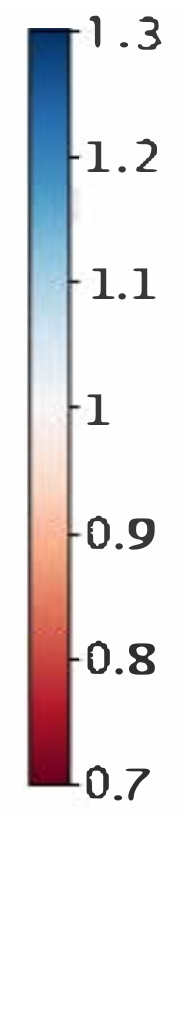}
%  \includegraphics[%trim=0 0.5in 0 0, clip,
%  width=\linewidth]{figs/_deg_4_legend_comb_rel.pdf}
\end{subfigure}%
%\vspace{-5.75mm}
\caption{Comparison between \emph{Chopin} and the \emph{centralized} scheduler under different traffic patterns (generated by scaling realistic flow size distributions and Poisson flow arrival times under different rates), when the optical ToR switch's connectivity is $4$. The color represents the ratio between the optical throughput of Chopin and the centralized~scheduler. 
%Blue cells mark settings where the distributed scheduler outperforms the centralized one; red cells mark the opposit The color represents Chopin's improvement (in percentage) compared to the centralized schedulers. 
As the blue cells become darker, Chopin more strongly outperforms the centralized scheduler.
}
\label{fig:mega_comb_graph_deg4_chopin-centralized}
\vspace{-4.5mm}
\end{figure*}

\begin{figure*}
%\vspace{-1.5mm}
\centering
\begin{subfigure}{0.31\linewidth}
  \centering
  \includegraphics[trim=0 0 1.0in 0in, clip, width=\linewidth]{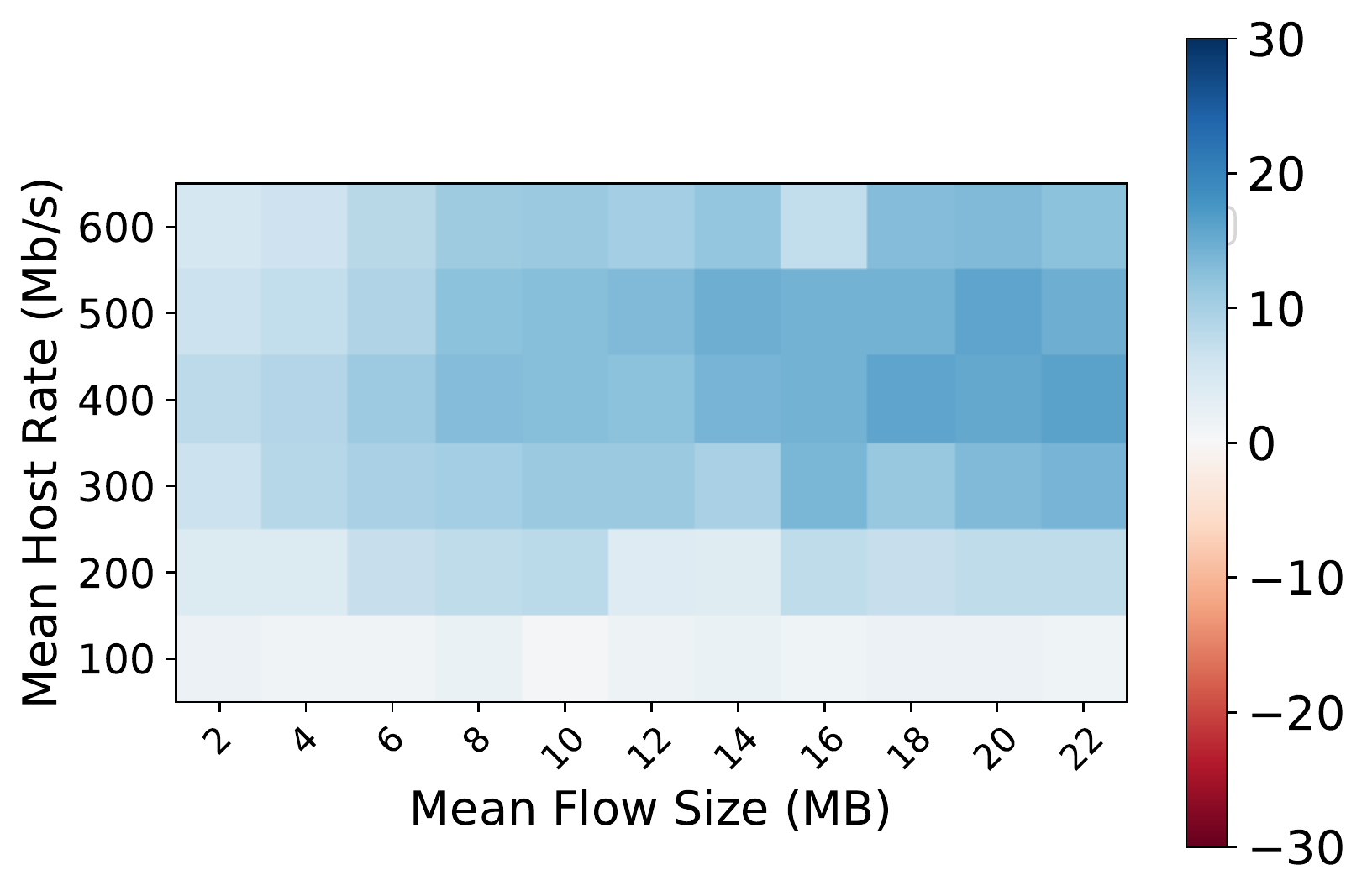}
 % \vspace{-7mm}
  \caption{HULL}%~\cite{HULL}}
  \label{fig:mega_comb_heatmap_deg4_chopin-dist_only_rel_HULL}
  \quad
\end{subfigure}%
\begin{subfigure}{0.31\linewidth}
  \centering
  \includegraphics[trim=0 0 1.0in 0, clip, width=\linewidth]{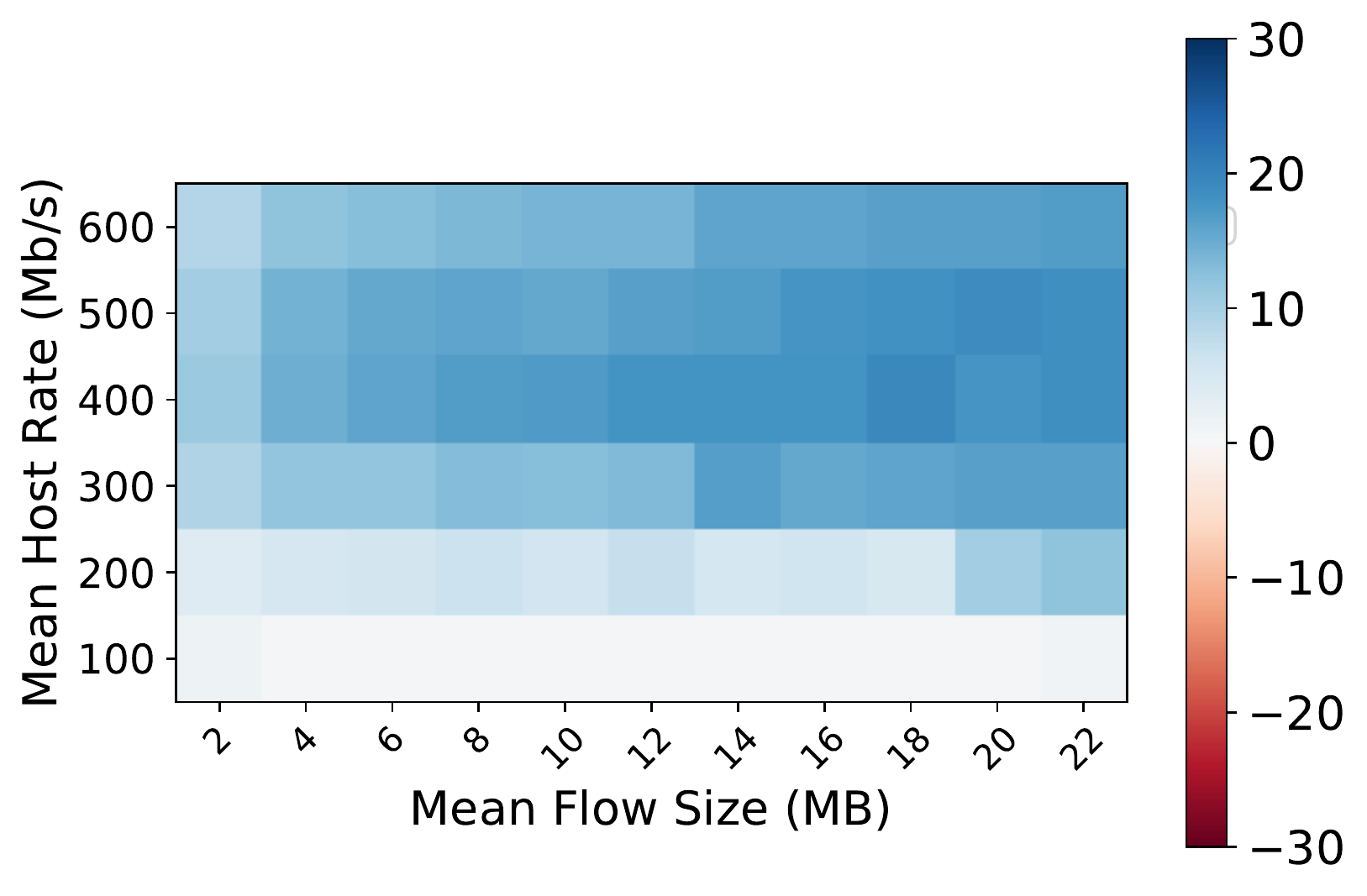}
  %\vspace{-7mm}
  \caption{pFabric}%~\cite{DCTCP}}
  \label{fig:mega_comb_heatmap_deg4_chopin-dist_only_rel_pFabric}
  \quad
\end{subfigure}%
\begin{subfigure}{0.31\linewidth}
  \centering
  \includegraphics[trim=0 0 1.0in 0, clip, width=\linewidth]{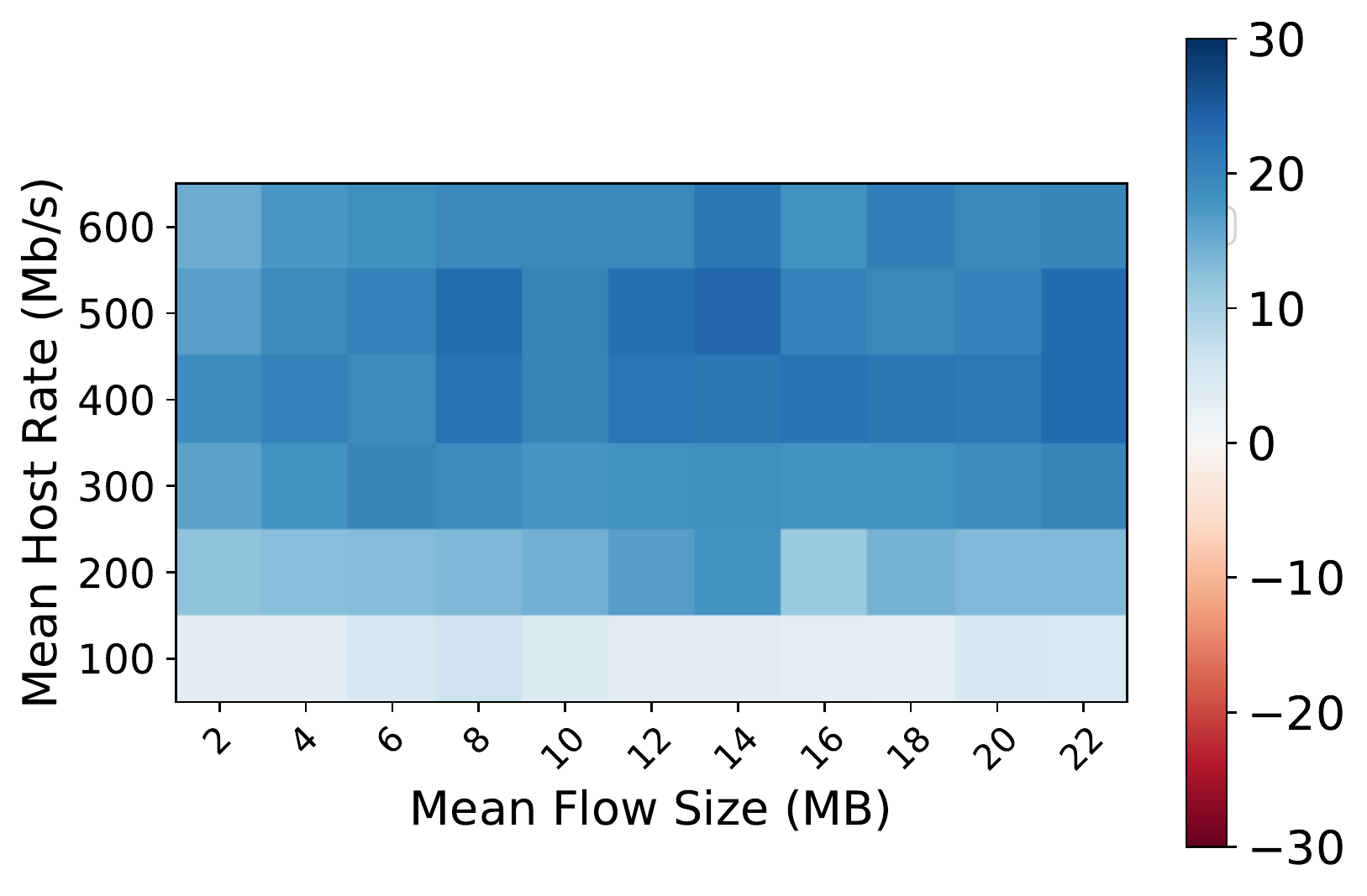}
  %\vspace{-7mm}
  \caption{VL2}%~\cite{VL2}}
  \label{fig:mega_comb_heatmap_deg4_chopin-dist_only_VL2}
  \quad
\end{subfigure}%
\begin{subfigure}{0.05\linewidth}
  \centering
  \includegraphics[%trim=0 0.5in 0 0, clip,
  scale=0.35]{figs/_deg_4_legend_comb_rel_david.pdf}
\end{subfigure}%
%\vspace{-5.75mm}
\caption{Comparison between \emph{Chopin} and the \emph{distributed} scheduler under the same settings as in Fig.~\ref{fig:mega_comb_graph_deg4_chopin-centralized}.
The color represents the ratio between the optical throughput of Chopin and the distributed scheduler. 
As the blue cells become darker, Chopin more strongly outperforms the distributed scheduler.}
\label{fig:mega_comb_graph_deg4_chopin-distributed}
%\vspace{-4.5mm}
\end{figure*}

\subsection{Scheduler Implementation}
\label{sec:evaluation_centralized}

The Chopin scheduler consists of centralized scheduler and distributed scheduler. The centralized,  
described in Section~\ref{sec:centralized}, aims to find a Maximum Weight Matching (MWM) solution. 

However, due to its complexity, especially as the optical degree ($k$) increases, \ignore{without further optimizations, }the centralized scheduler suffers from large running times. 
In order to reduce delays, two approximations were introduced. First, MWM with degree $k$ is computed as an iterative Edmond's MWM algorithm. Second, the centralized scheduler considers only the top-$m$ live flows per ToR switch (instead of all possible pairs between the $n$ nodes). We found top-$5$ MWM  running time to be within $1\%$ of the MWM over all pairs, since MWM complexity (which is the core of our b-matching solution) scales linearly with the number of to-be-matched edges.
%, this implies an $\Omega(n)$-factor reduction in computation time (e.g., for a datacenter with $80$ ToR switches, from up to $80\cdot(80-1)/2 = 3160$ to $80\cdot 5 = 400$). 
%
%
%includes one compute epoch delay %(i.e., ignoring computation and transfer~time).
%
%
%

%
Moreover, as each node reports only its top-$5$ nodes to the controller, the report can be sent by a single $200$ bit packet. Considering $100$ switches reporting to a controller with $1$Gbps network card, and control plane latency of $0.05$ ms, all reports can be sent within $0.07$ ms. The reconfiguration commands (for at most $4$ links per ToR switch) will have similar~latency.
%The computation is as follows:
%top-5 can be sent within 200 bit packets, and with 100 ToRs, 200*100/(10^6) = 20 microsec

Lastly, we consider the actual update time of the switch internal configuration after the reconfiguration message arrives. However, it is considered as negligible, assuming an optimized implementation with time complexity dominated by TCAM update time which is approximately in the $0.025$ ms range \cite{TCAM_update_latency}. Therefore, the total reconfiguration latency based on top-$5$ nodes can be bounded by $3$ ms.

%\cite{MWM_computation_complex}, I think this cite is problematic as it is on approximations and we might not want to open that up here
%this implies to $n/5$ reduction in computation time (e.g.,  factor of $16$ for a datacenter with $80$ ToRs).

%\subsection{Scheduler Evaluation Benchmarks}\label{subsec:sd}
\subsection{Scheduler Evaluation Benchmarks} 
We evaluate Chopin with respect to the following centralized and distributed schedulers. 

\vspace{0.1in}\noindent{\bf Centralized schedulers} are designed for long term datacenter flows.
The realistic centralized scheduler was analyzed through different values of the centralized scheduler epoch $T$, and with delay $\Delta$ equals to $T$.
Namely, in the third epoch, the scheduler uses matching results based on data collected in the first epoch, i.e., data from two epochs ago, recall Fig.~\ref{fig:centrelized_parameters} (characterized both \emph{centralized scheduler} and \emph{Chopin centralized scheduler}).
 Similarly to Veisllari et al.~\cite{Solstice}, we consider an \emph{optimal scheduler}, which runs MWM, \emph{with access to future traffic knowledge}. For each $1$ ms interval it uses the optimal matching computed as MWM of that interval. Therefore, it is an upper bound for datacenter performance.
 
We also consider an \emph{online optimal scheduler}, which has no knowledge of the future but it 
does not suffer from any delay. For each $1$ ms interval it uses an allocation computed as the MWM of the previous interval.

\vspace{0.1in}\noindent{\bf Distributed schedulers} are designed for bursts and short datacenter flows.
According to Roy et al.~\cite{Facebook_2015}, $90\%$ of the time, $50\%$ of the heavy flows change within $1$ ms. Therefore, the distributed scheduler should operate repeatedly in high frequency. \emph{Chopin's distributed scheduler} is set to operate every 1 ms, which is the length of a time-slot in our model. Furthermore, as in the centralized scheduler, both aggregation interval and delay ($a$ and $\delta$ respectively) are set to equal the time between two invocations (namely, $1$ ms). 
In addition, the performance of a \emph{distributed scheduler} (unrelated to a centralized scheduler) with the same properties was also analyzed.
Moreover, as discussed in \S\ref{sec:distributed}, a major factor of the Chopin distributed scheduler is the threshold $\alpha$, the level under which the centralized allocation can be changed by the distributed scheduler.

As the threshold decreases, Chopin's performance is closer to a centralized scheduler. Similarly, as the threshold increases, Chopin's performance is closer to being distributed. % scheduler.
Therefore, we evaluate Chopin for different threshold levels, between $0.1$ to $1.3$, to capture Chopin's performance scheduling between distributed and centralized scheduling.
%The \emph{Chopin scheduler} combines both Chopin's distributed scheduler and centralized scheduler, by threshold $\alpha$.

%\subsection{Centralized-Distributed Trade-off}\label{sec:Trade-off}
\subsection{Centralized-Distributed Trade-off} 
How can we find an optimal tradeoff between the centralized scheduler,
which provides accurate solutions but relies on outdated information,
and the distributed scheduler which relies on more recent information but provides approximate solutions (due to locality)?

This trade-off was analyzed in two related ways: (i) optimal threshold, and (ii) optimal reconfiguration number.
The threshold $\alpha$ is the parameter which enables the distributed scheduler to change the centralized matching, and therefore, to adapt the traffic changes in small time intervals. For example, a circuit allocation between ToR pair with high throughput on previous intervals, should be torn down if the flow rate reduces drastically.
We found that Chopin's optimal threshold $\alpha$ is between $\approx0.4-0.7$, depending on the traffic pattern. For the HULL traffic pattern, it achieves higher performance with $\alpha = 0.4$, while for DCTCP and VL$2$ traffic, the optimal threshold is approximately $0.7$.  
Moreover, across this range, the performance across all the traffic patterns were the highest, with low deviation.

\begin{figure*}%[htbp]
%\vspace{-5.75mm}
%\vspace{-6.5mm}
\hfill
\centering
\begin{subfigure}{0.29\linewidth}
  \centering
  \includegraphics[width=\linewidth]{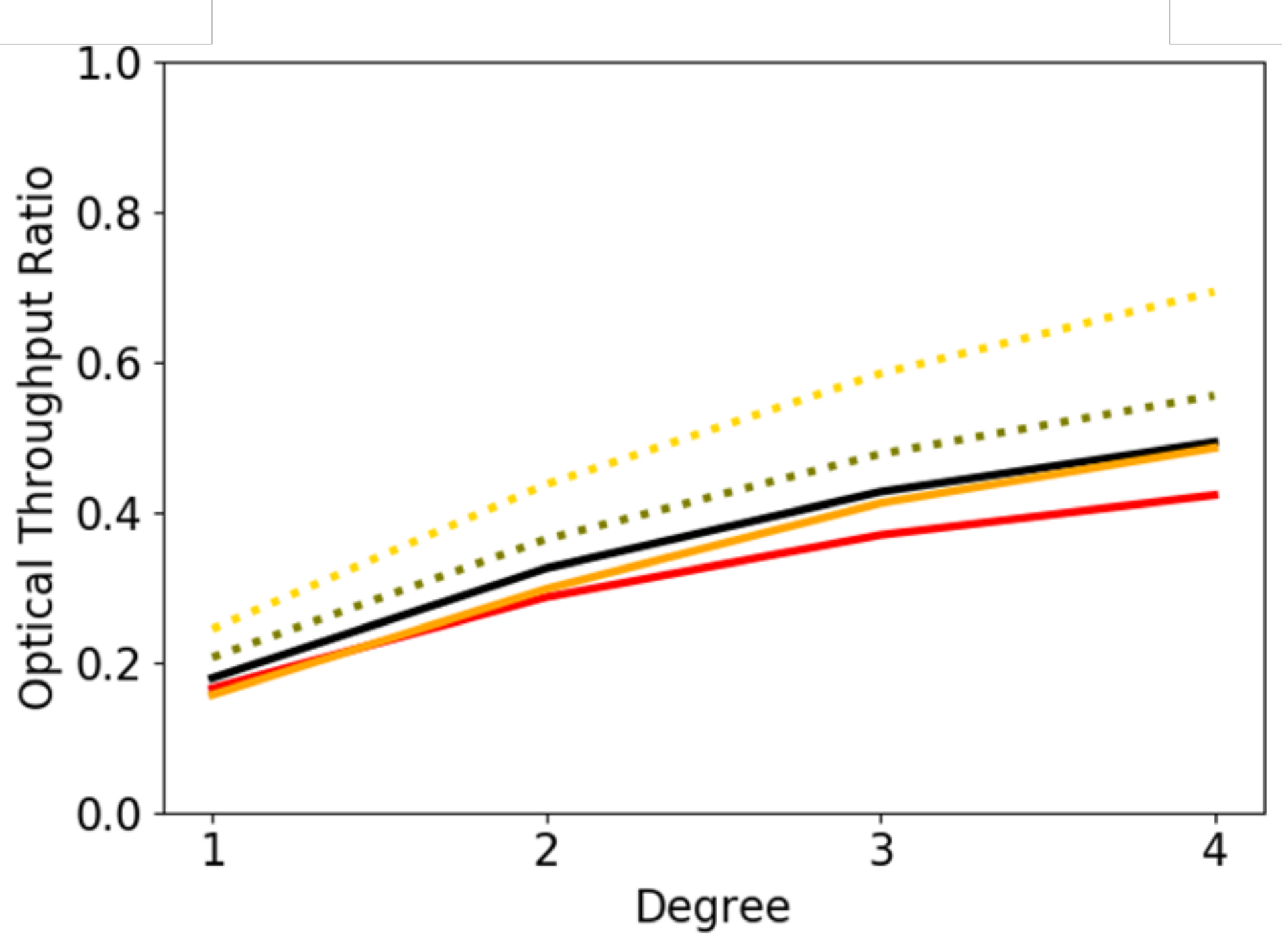}
  %\vspace{-6mm}
  \caption{HULL}
  \label{fig:HULL_deg_3}
  \quad
\end{subfigure}%
\begin{subfigure}{0.29\linewidth}
  \centering
  \includegraphics[width=\linewidth]{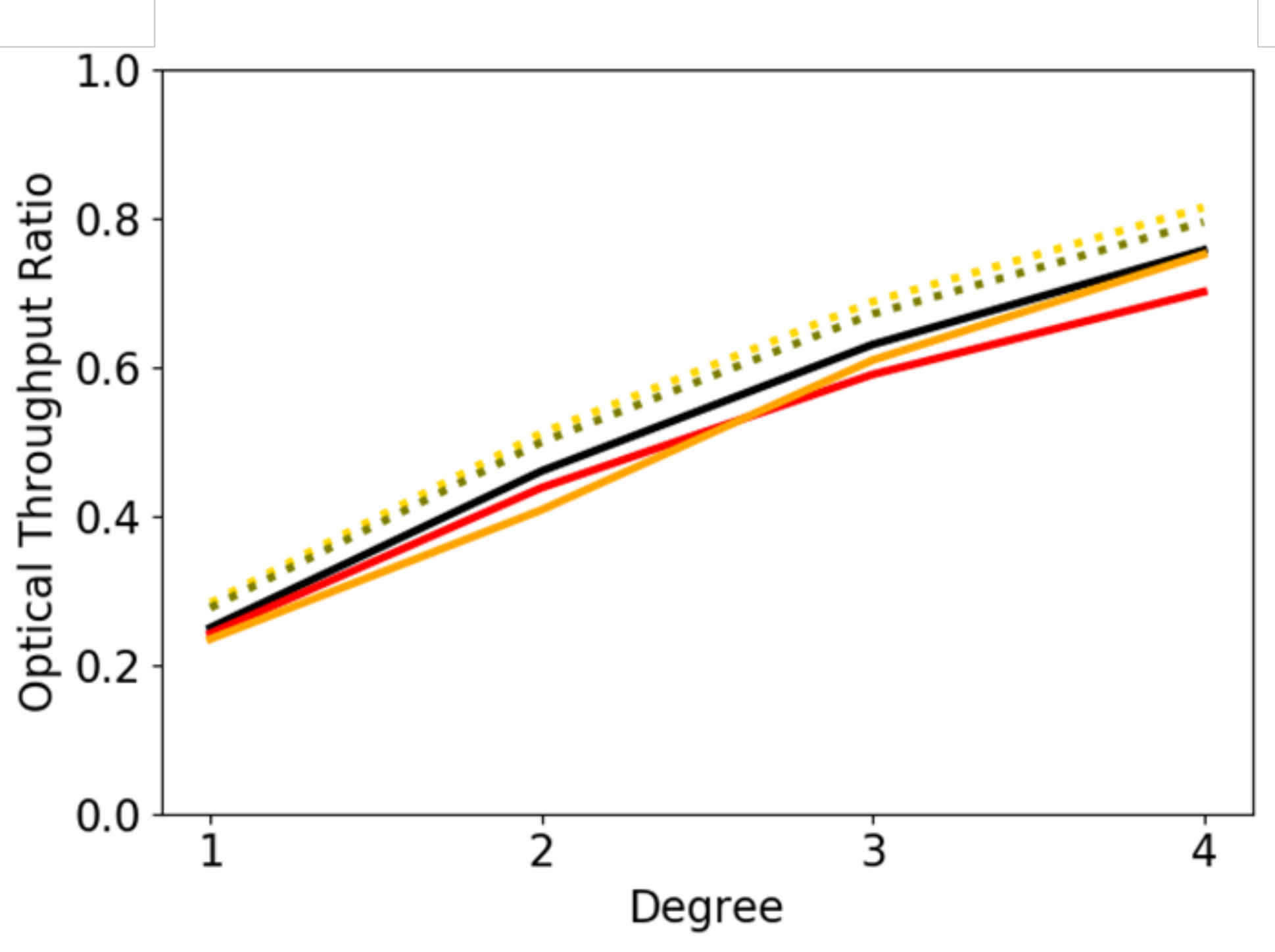}
  %\vspace{-6mm}
  \caption{pFabric}
  \label{fig:pFabric_deg_3}
  \quad
\end{subfigure}%
\begin{subfigure}{0.29\linewidth}
  \centering
  \includegraphics[width=\linewidth]{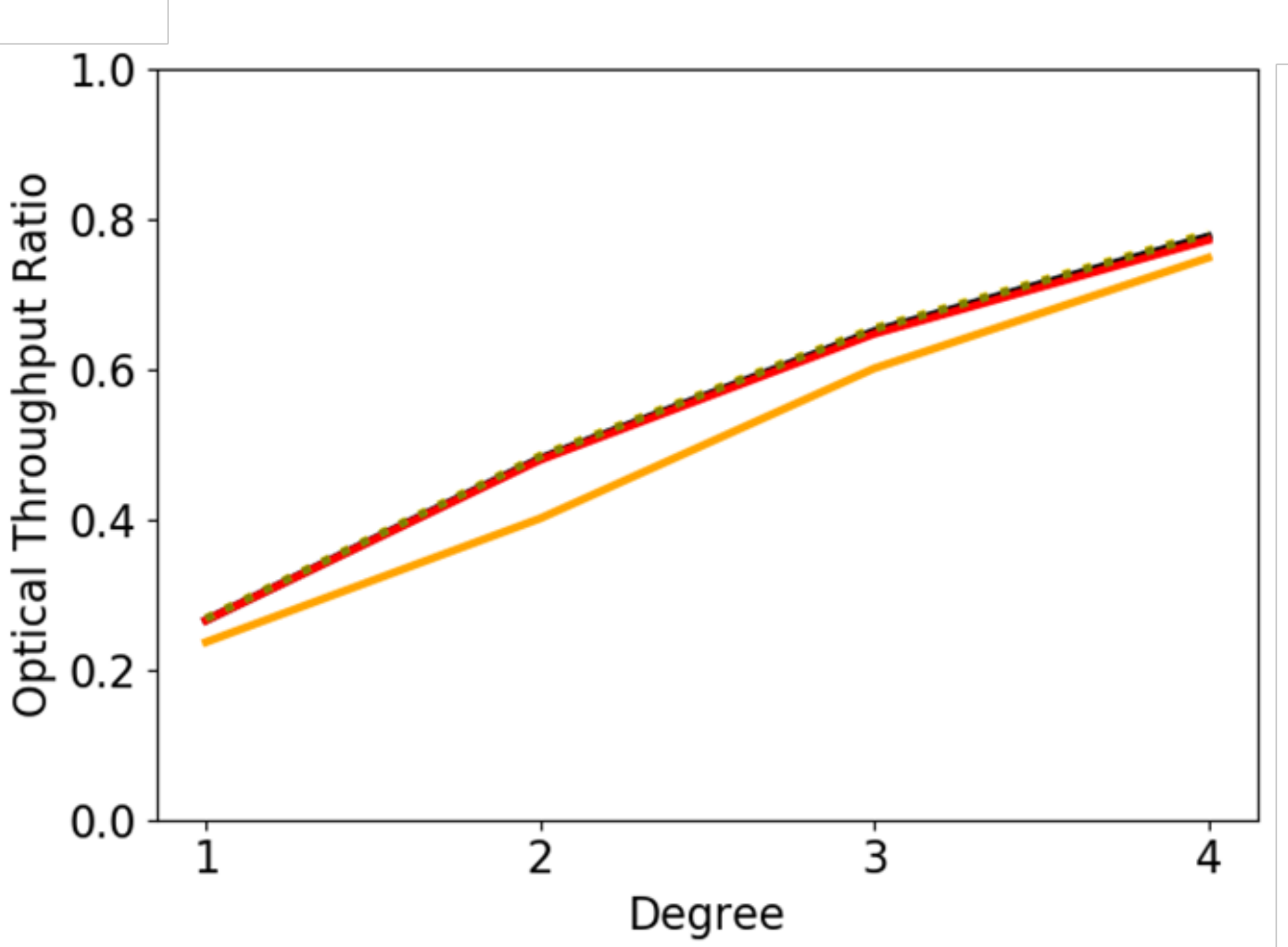}
  %\vspace{-6mm}
  \caption{VL2}
  \label{fig:VL2_deg_3}
  \quad
\end{subfigure}%
\begin{subfigure}{0.1\linewidth}
  \centering
  \includegraphics[width=\linewidth]{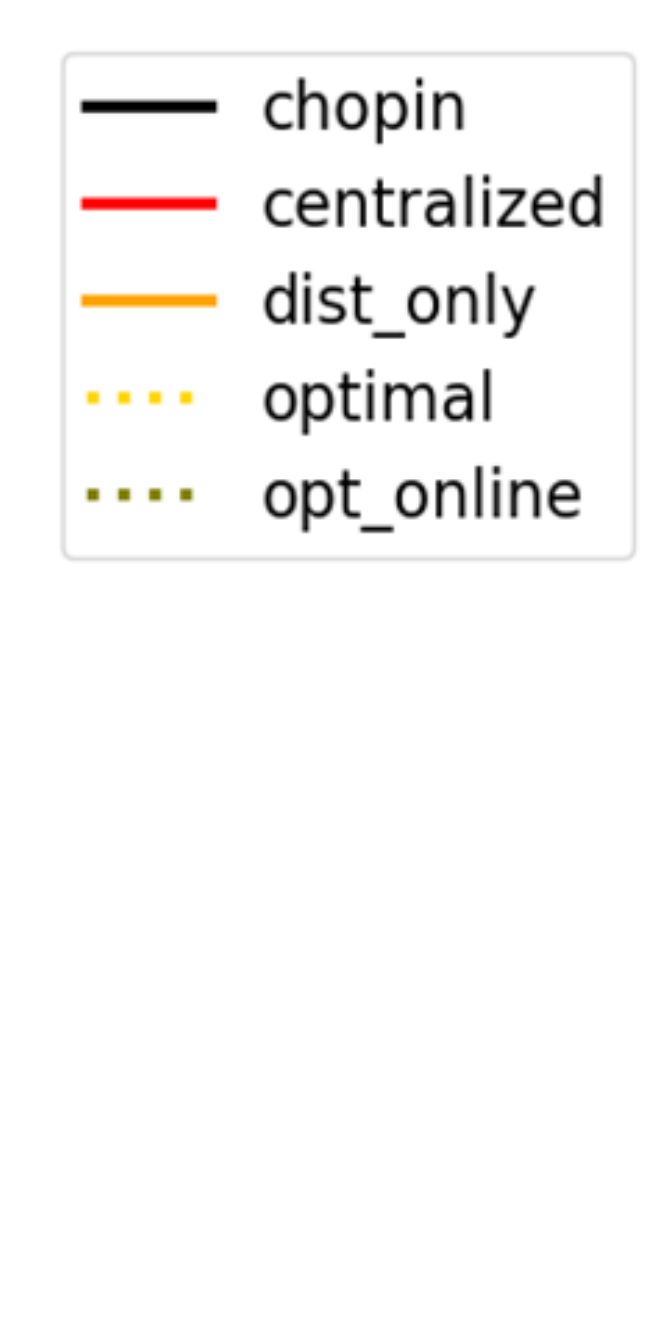}
\end{subfigure}%
%\vspace{-5.75mm}
\caption{Throughput through the optical circuits, for different optical degrees and flow size distributions.}
\label{fig:degrees_throughput}
%\vspace{-1mm}
\end{figure*}

\begin{figure*}%[htbp]
\vspace{-4.5mm}
\centering
\begin{subfigure}{0.30\linewidth}
  \centering
  \includegraphics[trim=0 0 1.0in 0, clip, width=\linewidth]{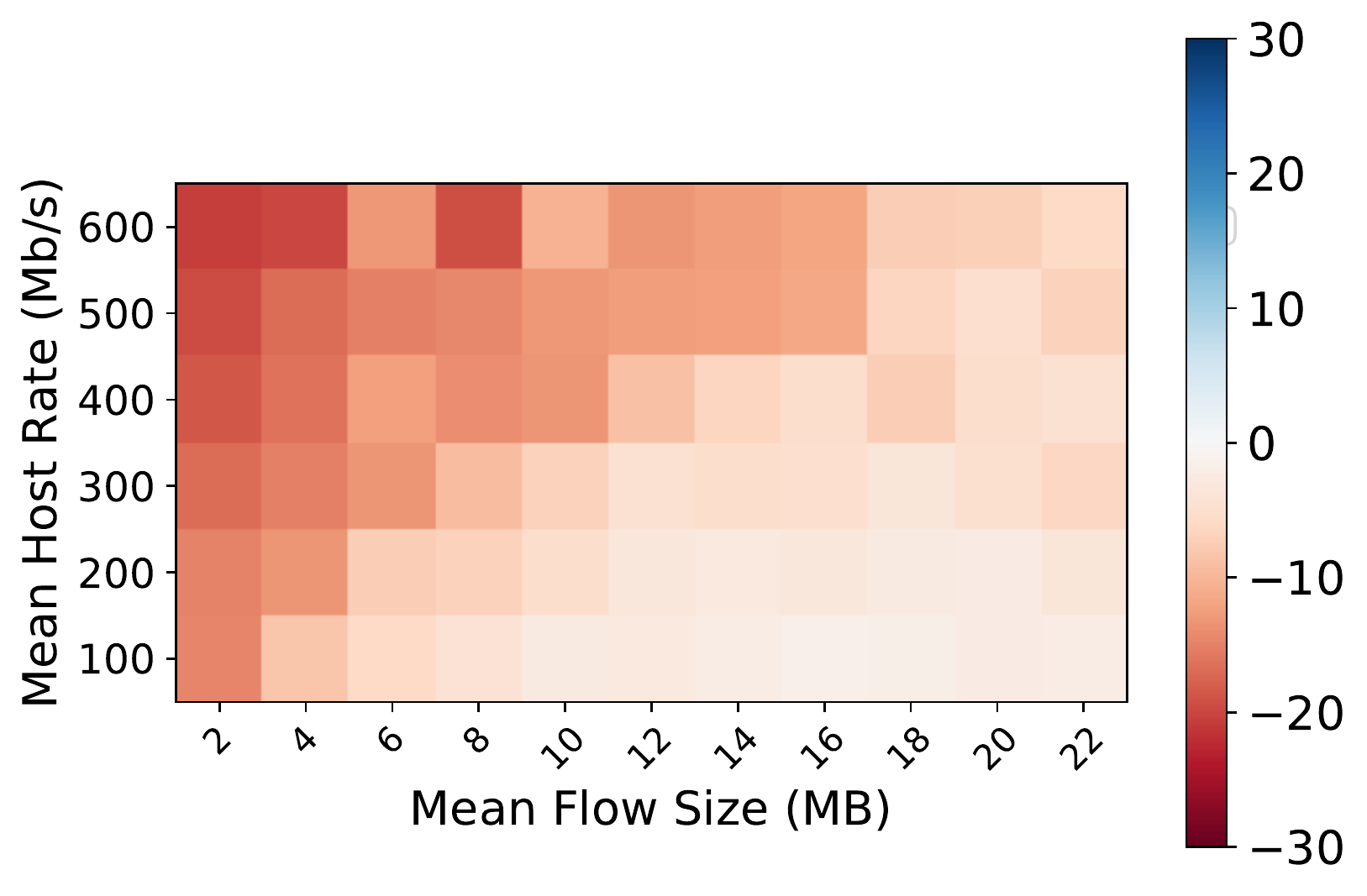}
  %\vspace{-7mm}
  \caption{HULL}
  \label{fig:mega_comb_heatmap_deg4_chopin-opt_online_rel_HULL}
  \quad
\end{subfigure}%
\begin{subfigure}{0.30\linewidth}
  \centering
  \includegraphics[trim=0 0 1.0in 0, clip, width=\linewidth]{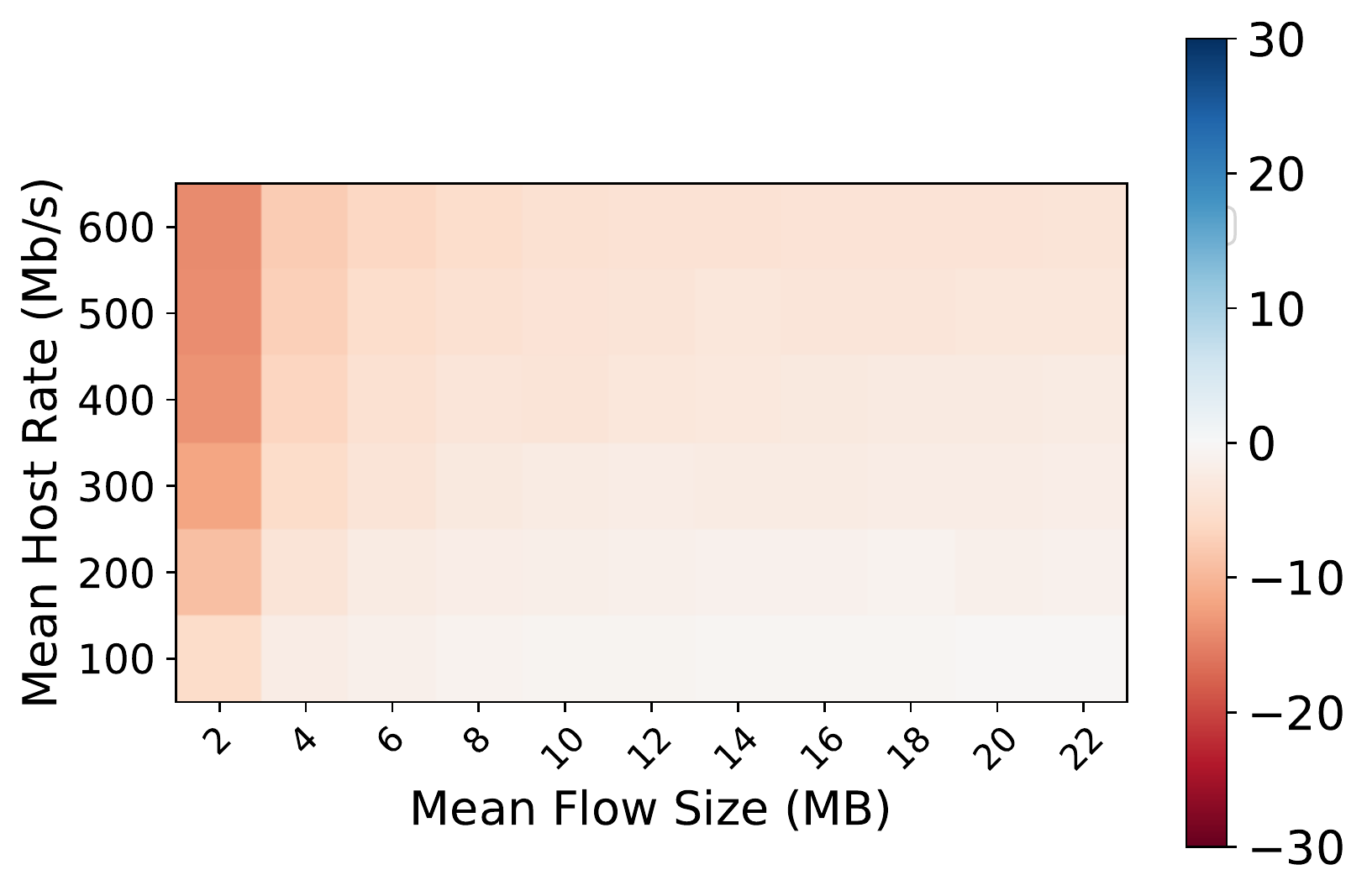}
 % \vspace{-7mm}
  \caption{pFabric}
  \label{fig:mega_comb_heatmap_deg4_chopin-opt_online_rel_pFabric}
  \quad
\end{subfigure}%
\begin{subfigure}{0.30\linewidth}
  \centering
  \includegraphics[trim=0 0 1.0in 0, clip, width=\linewidth]{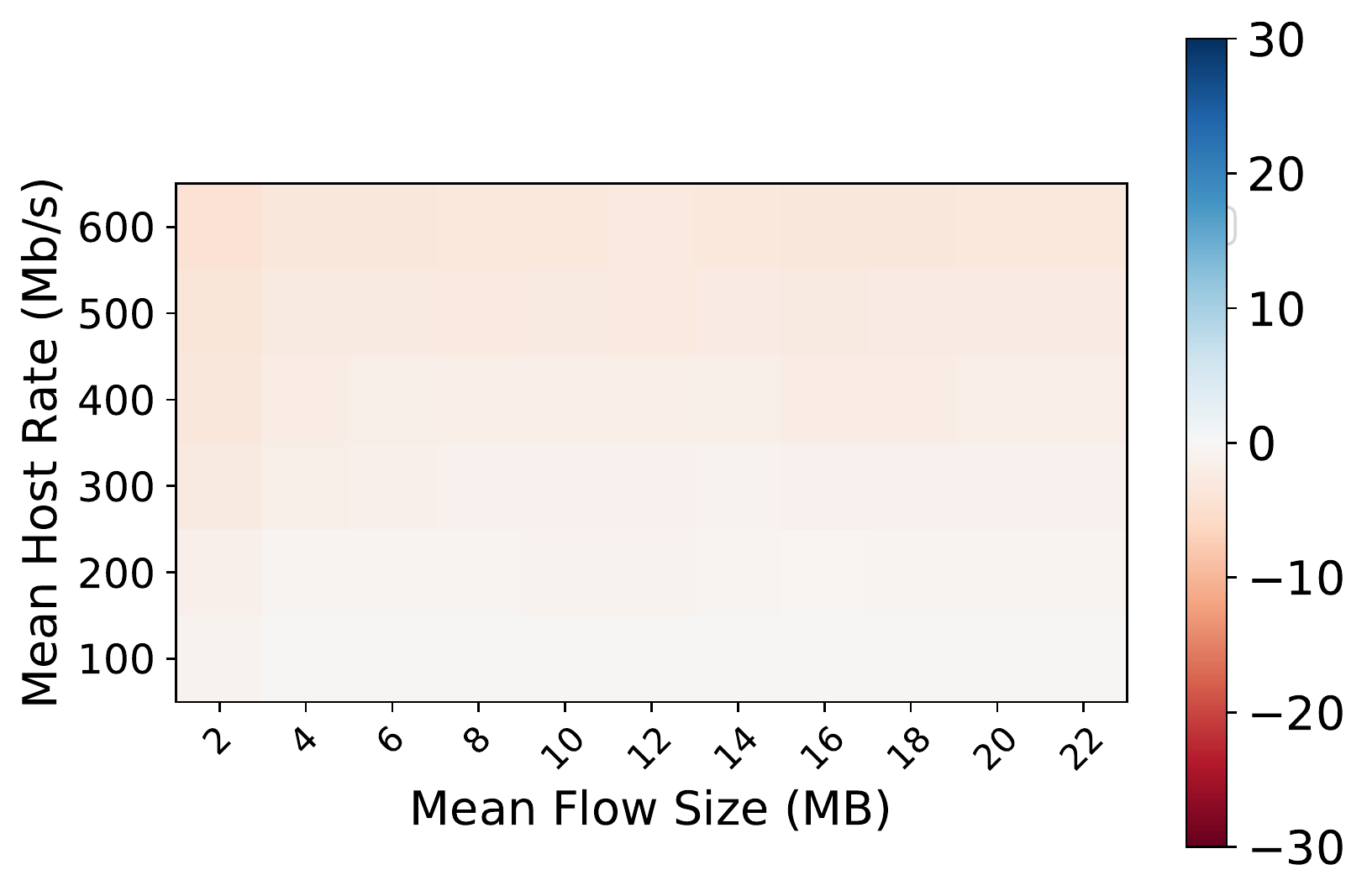}
 % \vspace{-7mm}
  \caption{VL2}
  \label{fig:mega_comb_heatmap_deg4_chopin-opt_online_rel_VL2}
  \quad
\end{subfigure}%
\begin{subfigure}{0.05\linewidth}
  \centering
  \includegraphics[%trim=0 0.5in 0 0, clip,
  scale=0.33]{figs/_deg_4_legend_comb_rel_david.pdf}
\end{subfigure}%
%\vspace{-5.75mm}
\caption{Comparison between \emph{Chopin} and the \emph{online optimal} scheduler under different traffic patterns (each generated by scaling well-known realistic flow size distributions and assuming Poisson flow arrival times under different rates), when the optical ToR switch's connectivity is $4$. As the red cells become darker, the online optimal scheduler performs better than Chopin.
}
\label{fig:mega_comb_graph_deg4_chopin-opt_online}
\vspace{-4mm}
\end{figure*}

\subsection{On the Benefit of Hybrid Scheduling}
%\vspace{-1mm}
To analyze Chopin's (the hybrid scheduler) improvement over distributed and centralized schedulers, we consider the optical throughput ratio. 
Fig.~\ref{fig:mega_comb_graph_deg4_chopin-centralized} and Fig.~\ref{fig:mega_comb_graph_deg4_chopin-distributed} describe Chopin's improvement ratio for each of the flow patterns, compared to centralized and distributed schedulers (respectively), when considering a centralized compute epoch of $3$ ms, see \S\ref{sec:evaluation_centralized}.

The results show that Chopin outperforms the centralized scheduler for \emph{every} traffic pattern. 
We found that Chopin's optical throughput ratio is higher than the centralized scheduler optical throughput ratio by up to $20\%$ in the HULL distribution, $15\%$ for pFabric pattern, and $2\%$ for datacenters with a VL$2$ flow size distribution (as shown in Fig.~\ref{fig:mega_comb_heatmap_deg4_chopin-centralized_rel_HULL}). Moreover, Chopin also achieves a higher throughput ratio compared to the distributed scheduler across all traffic patterns. Specifically, Chopin increases the optical throughput ratio of the distributed scheduler by up to $16\%$ in the HULL distribution, $20\%$ in the pFabric and $23\%$ in VL$2$ traffic (see Fig.~\ref{fig:mega_comb_heatmap_deg4_chopin-dist_only_rel_pFabric}). Therefore, Chopin outperforms both centralized and distributed schedulers.

\subsection{Optical Degree Improvement}
%\vspace{-1mm}
Next, we examine the improvement as a function of Chopin nodes' optical degree. Therefore, we have focused on the optimal\ignore{maximal} threshold  for  each  of  the  flow  patterns,  where  the centralized scheduler epoch is 3 ms, as discussed in Section~\ref{sec:centralized}.

Fig.~\ref{fig:degrees_throughput} presents the  ratio between the throughput through optical circuits to the overall throughput (electrical and optical networks combined). This \emph{optical throughput ratio} changes with the optical degree and flow patterns, as shown in Fig.~\ref{fig:degrees_throughput}. In each flow pattern, all the schedulers were considered. 
It is shown that as the degree increases, the throughput among all the schedulers improved, and that Chopin's throughput is higher than both the centralized and the distributed schedulers. Moreover, as the number of elephant flows increases (as in VL$2$), Chopin's throughput is getting closer to the optimal. It is consistent with Chopin's aim to carry elephant flows over optical circuits. Therefore, flow patterns with high number of elephant flows benefit more from using Chopin.

\subsection{Chopin VS Online Optimal Scheduler}

We compared between Chopin performance, and online optimal scheduler performance (where centralized updates are being sent to Chopin nodes every $1$ ms instead of $3$ ms respectively). Fig.~\ref{fig:mega_comb_graph_deg4_chopin-opt_online} 
shows that even if Chopin's centralized scheduler updates were sent every $1$ ms (such as in the online optimal scheduler), there is no significant improvement for the $VL2$ flow pattern (see Fig.~\ref{fig:mega_comb_heatmap_deg4_chopin-opt_online_rel_VL2}). In other words, Chopin is closer to the optimal scheduler as the flows become larger, because as the mean flow is longer, the changes over small time intervals (such as $1$ ms) become minor. Therefore, in these cases, the added value of high frequent scheduling updates, even with the ``future'' information (as in the optimal scheduler), decreases. Chopin can benefit from higher frequent centralized updates mostly in HULL distribution, (where the flows are usually shorter), by approximately $20\%$.

%\balance

%\begin{wrapfigure}{r}{0.5\textwidth}
\begin{figure}[t]
%\vspace{-6mm}
  \begin{center}
      \begin{subfigure}[t]{0.33\linewidth}
  \centering
  \includegraphics[%trim=0 0 1.0in 0, clip, 
  width=\linewidth]{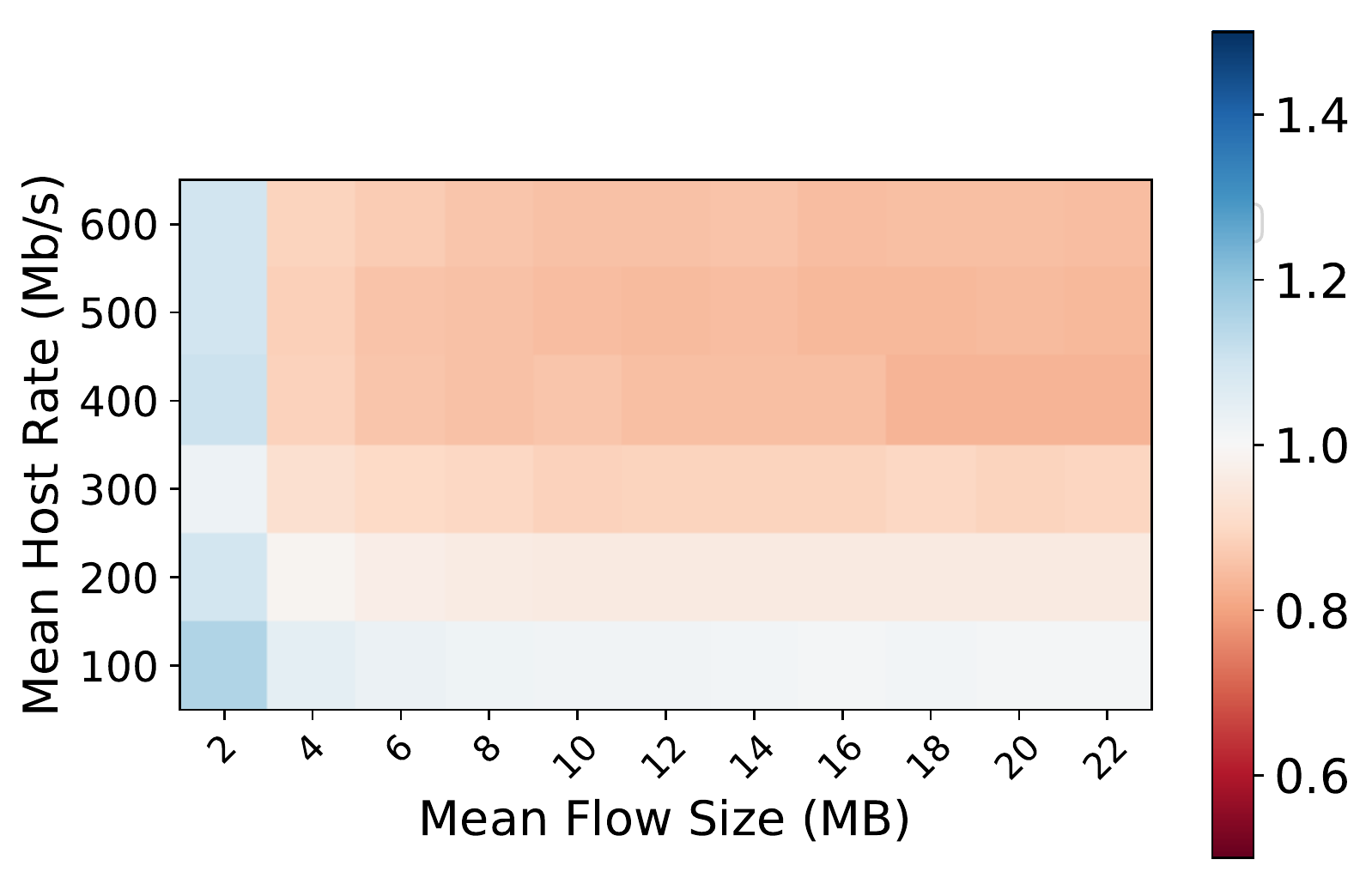}
\end{subfigure}%
\begin{subfigure}[t]{0.1\linewidth}
  \centering
  \includegraphics[%trim=0 0.5in 0 0, clip, width=\linewidth,
  scale=0.275]{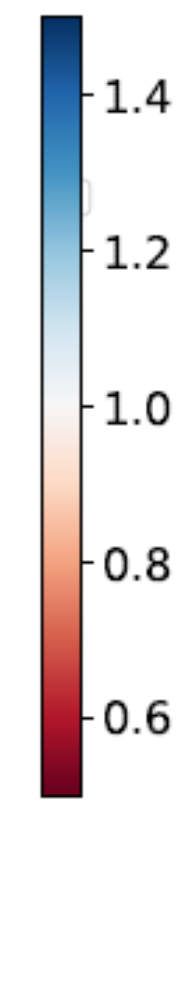}
\end{subfigure}  
  \end{center}
 % \vspace{-5mm}
  \caption{\label{fig:comb_deg_4_reconf_160_pFabric} Comparison between centralized and distributed schedulers, as in Fig.~\ref{fig:mega_comb_graph_deg4}, but with $160$ ToR switches. For brevity we only present pFabric results.}
%  \vspace{-3mm}
\end{figure}
%\end{wrapfigure}

\subsection{Sensitivity Analysis}\label{subsec:sensa}
We further analyzed our results by considering different sizes of networks, e.g.\ with $100$ ToR switches and for $160$ ToR switches, where there are $10$ hosts per ToR switch, and for different rates per host ($100$--$600$ Mbps).
Across all the networks that were examined, for each of the traffic patterns, we observe that on certain conditions the distributed scheduler outperforms the centralized scheduler and vice versa, with respect to  higher optical throughput. For instance, see the heatmap for a network with $160$ ToR switches, each with an optical degree of $4$ connections, under the pFabric traffic pattern in Fig.~\ref{fig:comb_deg_4_reconf_160_pFabric}. It shows that under the pFabric distribution, when the mean flow is larger than $5$ MB, the centralized scheduler achieves higher performance compared to the distributed one. This phenomenon was also observed for the $80$ ToR network, as in  Fig.~\ref{fig:mega_comb_heatmap_deg4_pfabric}.

\begin{comment}
\begin{figure}%[tbp]
  \centering
  \begin{subfigure}[t]{0.42\linewidth}
  \centering
  \includegraphics[%trim=0 0 1.0in 0, clip, 
  width=\linewidth]{figs/mega_comb_heatmap_deg4_dist_only-centralized_rel_pFabric_160_david.pdf}
\end{subfigure}%
\begin{subfigure}[t]{0.1\linewidth}
  \centering
  \includegraphics[%trim=0 0.5in 0 0, clip, width=\linewidth,
  scale=0.35]{figs/dist_cent_deg_4_legend_comb_rel_2.pdf}
\end{subfigure}  
%\vspace{-3.75mm}
\caption{\label{fig:comb_deg_4_reconf_160_pFabric} Comparison between centralized and distributed schedulers, as in Fig.~\ref{fig:mega_comb_graph_deg4}, but with $160$ ToR switches. For brevity we only present pFabric results.}
%\vspace{-5mm}
\end{figure}
\end{comment}

Moreover, Chopin’s performance can scale. We demonstrate its effectiveness over a concrete network topology (specified in Section~\ref{sec:Implementation}), but faster links with higher demand will create the same bottleneck and resolve with Chopin in the same way.

%\balance

\section{Conclusion}
\label{sec:conclusions}
%
%\textbf{Chopin: centralized \emph{and} distributed scheduling.}
Chopin aims to combine the benefits of centralized scheduling with distributed scheduling, to provide high throughput and fast reaction.
While centralized and distributed scheduling has also been combined in all-static non-hybrid networks, e.g., Facebook's Express Backbone~\cite{fbcd}, hybrid networks with optical circuits pose structurally different challenges. \ignore{ by ongoing reconfiguration.}
In particular, we find that distributed decisions benefit from being closer in time to the measurements they are based on, which is more critical than the rate of decisions. 

%\section{Future Research}

We believe that our work opens several interesting avenues for future research. 
In particular, while we achieve significant performance gains, our approach is more complex than the state-of-the-art and it would be useful to simplify it further. Furthermore, our distributed schedulers use the same threshold for all nodes as a homogeneous strategy.
While this succinct representation is sufficient for the settings described in this paper, it can be interesting to explore heterogeneity, e.g., to increase the threshold on very congested racks.
%at the nodes' parameters (e.g., to increase the threshold on very congested racks). 
%
Finally, the trade-off between 
%a ToR pair 
an elephant flow's duration and the time before it starts to route through optical circuits
%Chopin can be considered for further utilization.
can be considered for future optimization.
%
%\clearpage

\begin{comment}
%\todo{maybe cut below, discuss?}
Finally, it would be interesting to see how Chopin technology can be used to enhance further scenarios.
This includes different topologies of the electrically switched network as well as different topologies of the optical network. To this end, we have looked on a scenario where the optical network is an optical ring (e.g., as illustrated in Fig.~\ref{fig:Chopin_Concept_3d}), as in Mordia~\cite{Mordia}. Such architectures often natively support multicast mode, where a single transmitter sends data to multiple or all receivers. In this way, we have a dedicated control circuit over the optical ring, in which ToR switches (or more specifically Chopin node) can quickly share information about their status and other optical circuit usage (e.g., in a TDM manner). Up until now, we have investigated how such information can enhance the performance of our distributed scheduler, where an initial study shows %a significant 
promising improvements. 
\end{comment}

%\paragraph*{Acknowledgments}
%We thank Liron Schiff for his comments, suggestions, and invaluable insights.

%\clearpage

\bibliography{Chopin-arxiv-2022}

%\clearpage

%\appendix
%\section{Scheduler Notation}
%\vspace{-10mm}
%\squeezeup

\ignore{
Table~\ref{table:notations} summarizes Chopin schedulers' notations.
\begin{table}[h]
\centering
\scriptsize
%\resizebox{\columnwidth}{!}{
\begin{tabular}{p{1.3cm} p{9.5cm}}
%\hline
\specialrule{.1em}{.05em}{.05em} 
Notation & Meaning  \\
\hline
%\hline
$K$ & Optical degree, the number of available circuits per Chopin node.\\ 
%\hline
$\Delta$ &  Centralized scheduler delay\\
%\hline
$\delta$ &  Distributed scheduler delay\\
 %\hline
$\alpha$ &  Chopin threshold for keeping centralized decisions \\
%\hline
$A$ &  Centralized scheduler aggregation interval \\
%\hline
$a$ &  Distributed scheduler aggregation interval \\
%\hline
$T$ & Centralized scheduler epoch\\
%\hline
\specialrule{.1em}{.05em}{.05em} 
\end{tabular}
%}
\caption{Chopin's \ignore{centralized and distributed} scheduler notation}
\label{table:notations}
%\vspace{-8mm}
\end{table}
}

%\scalebox{0.5}{
%\begin{minipage}{0.7\linewidth}
%\clearpage

\appendix
\section{Chopin's Distributed Scheduler Algorithm}\label{appendix:alg}
We provide on the next page the pseudo-code of Chopin's distributed algorithm in Algorithm~\ref{alg:distributed_alg}.
\begin{algorithm}[h]
	\caption{Chopin Distributed Algorithm Code for Node $i$.}
	\label{alg:distributed_alg}
	\begin{algorithmic}[1]
	\small
	%\Require current optical destination ToR $cur\_d$,
	\Statex max\_reqs : The number of allowed requests per ToR switch
    \Statex cur\_nodes : The nodes currently connected with $i$
\Statex centralized\_nodes : The nodes matched to $i$ by the centralized scheduler in its last invocation

\Statex received\_reqs $\leftarrow \emptyset$
\Statex \ 

\vspace{-3mm}

\Statex \hspace{-2em} Upon the beginning of a distributed scheduler epoch:

\Function{start:}{}
\State matched\_nodes $\leftarrow \emptyset$
\For {$p \in$ (cur\_nodes $\cap$ centralized\_nodes)}
    \If {$r_{i,p} \geq \alpha\cdot R_{i,p}$}
        \State matched\_nodes.add(p)
    \EndIf
\EndFor

\State req\_nodes $\leftarrow \left([n] \setminus \{i\}\right) \setminus
\mbox{matched\_nodes}$ 
\Comment $n$ denotes the \Statex \hfill number of Chopin nodes in the network
\State req\_nodes $\leftarrow$ {\sc get\_top\_nodes}(req\_nodes,max\_reqs)
\Statex \Comment Top max\_reqs nodes, out of req\_nodes, with  \Statex \hfill  the most bi-directional traffic with ToR switch $i$.
%\State granted = $\bot$ 
\State grants  $\leftarrow \emptyset$; denies $\leftarrow\emptyset$
%\State sent\_grants = False
%\State ended = False
\State {\sc send\_requests}(req\_nodes) \Comment Send \texttt{request} to all
\Statex \hfill   nodes in req\_nodes.
\EndFunction
\Statex \

\vspace{-4mm}

\Statex\hspace{-2em} Upon receiving a \texttt{request} message from src\_id:
\Function{request\_handler}{src\_id}:
\State received\_reqs.add(src\_id)
\EndFunction
\Statex \

\vspace{-4mm}

\Statex\hspace{-2em} Upon a timeout event (implying the request phase has ended):
\Function{request\_timeout\_handler}{}:
%\State $received\_reqs = get\_received\_reqs()$
\State nodes $\leftarrow$ req\_nodes $\cap$ received\_reqs
\State free\_links $\leftarrow$ k - $|$matched\_nodes$|$
\State granted $\leftarrow$ {\sc get\_top\_nodes}(nodes, free\_links)
%\State $new\_peers = \argmax\limits_{S \subseteq new\_peers, |S|\leq free\_links} \sum_{j\in S} r_{i,j}$
\State rejected $\leftarrow$ received\_reqs  $\setminus$ granted
\State {\sc send\_denies}(rejected) \Comment Send \texttt{deny} message to all \Statex \hfill nodes in rejected set.
\State {\sc send\_grants}(granted) \Comment Send \texttt{grant} message to  \Statex \hfill all nodes in granted set.
\State grant\_sent $\leftarrow$ true
\State {\sc try\_execute\_decisions}()
\EndFunction
\Statex \ 

\vspace{-4mm}

\Statex \hspace{-2em}Upon receiving a \texttt{grant} message from src\_id:
\Function{grant\_handler}{src\_id}:
        \State grants.add(src\_id)
        \State {\sc try\_execute\_decisions}()
%\If{denies $\cup$ grants == granted}
%\State {\sc execute\_decisions}()
%\EndIf
\EndFunction

\Statex \ 

\vspace{-3mm}

\Statex\hspace{-2em}Upon receiving a \texttt{deny} message from src\_id:
\Function{deny\_handler}{src\_id}:
        \State denies.add(src\_id)
                \State {\sc try\_execute\_decisions}()

%\If{denies $\cup$ grants == granted}
%\State {\sc execute\_decisions}()
%\EndIf
\EndFunction

%\algstore{myalg}
%\end{algorithmic}
%\end{algorithm}
%\begin{algorithm}
    
%\begin{algorithmic}
%\algrestore{myalg}

\vspace{-3mm}

\Statex \ 
\Function{try\_execute\_decisions}{}:
%\Comment Invoked when all \Statex \hfill \texttt{grant}/\texttt{deny} messages are received
%\If {not CompareAndSet(\&ended, False, True)}
%\State \Return
%\EndIf
\If{denies$\cup$grants$\neq$ req\_nodes {\bf or not} grant\_sent}
\State {\bf return} \Comment Not all grant/deny were received
\EndIf
%\Statex \hspace{2em}$\triangleright$ Grant phase has ended
\State new\_nodes$\leftarrow$granted $\cap$ grants
\For {p$\in$(cur\_nodes$\setminus$ new\_nodes)$\setminus$matched\_nodes}
    \State {\sc disconnect}(p)
\EndFor
\For {p $\in$(new\_nodes$\setminus$cur\_nodes$\setminus$ matched\_nodes)}
    \State {\sc connect}(p)
\EndFor
\State received\_reqs $\leftarrow\emptyset$; grant\_sent$\leftarrow$false
\EndFunction
\normalsize
\end{algorithmic}
\end{algorithm}
%\vspace{-3.5mm}
%\end{minipage}

\end{document}